\documentclass[12pt]{article}

                              \usepackage {subeqn,cite,psfig}
\def\be{\begin{equation}}
\def\ee{\end{equation}}
                              \def\barr{\begin{array}}
                              \def\earr{\end{array}}
\def\dis{\displaystyle}

\def\etal{{\em et al.}}

\def\lsim{\:\raisebox{-0.5ex}{$\stackrel{\textstyle<}{\sim}$}\:}
\def\gsim{\:\raisebox{-0.5ex}{$\stackrel{\textstyle>}{\sim}$}\:}
                              
                              \def\gev{\: {\rm GeV} }
                              \def\tev{\: {\rm TeV} }
                              
                              \def\fb{\: {\rm fb}}
\def\ra{\rightarrow}

\def \rp {$R_p \hspace{-1em}/\;\:$}

\newcommand{\beq}{\vspace{2mm}\begin{eqnarray}}
\newcommand{\eeq}{\end{eqnarray}\vspace{2mm}}

\def\l{\lambda}
\def\lp{\lambda'}
\def\snu{\widetilde \nu}

\begin{document}
\setcounter{page}{0}
\thispagestyle{empty}
\setcounter{footnote}{0}
\renewcommand{\thefootnote}{\fnsymbol{footnote}}
\begin{flushright}
                                             CERN-TH/98-222\\[1.5ex]
                                             MRI-PHY/P980756\\[1.5ex]
                                      {\large\sf  hep-ph/9807373} \\
\end{flushright}
\vskip 25pt
\begin{center}
\advance\baselineskip by 10pt

{\Large\bf Signals of $R$-Parity Violation at a Muon Collider        }
\\[2cm]

\advance\baselineskip by -10pt

{\bf                    Debajyoti Choudhury                          }

{\footnotesize\em Mehta Research Institute 
     for Mathematics and Mathematical Physics, \\
     Chhatnag Road, Jhusi, Allahabad -- 211 019, India. \\
     Electronic address: debchou@mail.cern.ch                 }\\[2ex]

\rm
\vspace{13pt}

{\bf 
                      Sreerup Raychaudhuri
}

{\footnotesize\em  Theory Division, CERN, CH 1211 Geneva 23, 
                   Switzerland. \\
                   Electronic address: sreerup@mail.cern.ch     }\\[2ex]

\vspace{30pt}

{\bf                       Abstract                              }
\end{center}

\begin{quotation}
We investigate the effects of $R$-parity-violating $LL\bar E$ and 
$LQ\bar D$ operators on the production of fermion pairs at a 
$\mu^+ \mu^-$ collider with high energy and luminosity. We show that 
comparison of
angular distributions leads to useful discovery limits for the
$R$-parity-violating couplings, especially if the exchanged 
sfermions turn out to be heavy. 
Products of these couplings can also be probed well beyond
present experimental limits, especially for the second and third
generations. Finally, we compare results at a muon collider with those
obtainable from an $e^+ e^-$ linear collider with similar energy and
luminosity.
\end{quotation}

\vskip 30pt

\begin{flushleft}
July 1998 \\
CERN-TH/98-222
\end{flushleft}

\setcounter{footnote}{0}
\renewcommand{\thefootnote}{\arabic{footnote}}
\newpage
\section{$\!\!\!\!\!\!.$~~Introduction}

The remarkable success of the Standard Model (SM) of strong and
electroweak interactions in explaining a wide range of phenomena at
various energies~\cite{SM_revs} has proved at once a triumph and an
embarrassment for high-energy physicists. The triumph is obvious, but the
embarrassment is there nevertheless, since the SM is only half-a-theory,
having a large number of {\em ad hoc} assumptions and phenomenological
inputs. Moreover, it suffers from a serious theoretical defect --- the
hierarchy problem --- which can only be circumvented by assuming that
there is some new physics beyond it at energy scales of a few TeV to a
few hundred TeV.

Supersymmetry, allowing the mixture of bosonic with fermionic degrees of
freedom, is perhaps the most popular and promising theoretical scenario
for physics beyond the SM, as it provides a natural solution for the
hierarchy problem. The Minimal Supersymmetric Standard Model
(MSSM)~\cite{mssm} is obtained from the SM in a straightforward way by
writing down the simplest supersymmetric model that contains the SM as a
subset. This involves inclusion of a second Higgs doublet for anomaly 
cancellation and some soft supersymmetry-breaking terms to account for 
the fact that no
supersymmetric partners of the known SM particles have been discovered
(yet), from which it must be inferred that they are heavy. The discovery
of supersymmetry will thus require higher energies and luminosities than
have been available till the present day. The quest for higher energies
has led to several ideas about new colliders and collider techniques, of
which one possibility has attracted a great deal of attention: 
$\mu^+ \mu^-$ colliders.

\section{$\!\!\!\!\!\!.$~~Muon Colliders.}

Current designs for high-energy colliders make use of beams of electrons,
protons or heavy ions. The latter can be accelerated to higher
energies, but these energies cannot be tuned in the same way as electron beam
energies can. Moreover, QCD backgrounds at hadron colliders are
often unmanageably large and render precise measurements difficult. The
great advantage of an $e^+e^-$ collider lies in the fact that the final
states are relatively clean and free of QCD backgrounds.

The basic motivation for a muon collider stems from the energy and
luminosity limitations of $e^+e^-$ colliders. The high luminosities
available at the currently running $e^+e^-$ facilities are achieved by
having multiple bunch crossings (typically a few thousand) in a storage
ring. However, storage rings have a well-known energy limitation due to
synchrotron radiation emitted by the rotating electron/positron beams,
which has rates inversely proportional to the fourth power of the mass of
the charged particle. As a result of this, a stage is inevitably 
reached when all the
energy pumped into the electron/positron beam is radiated away. Thus, it
seems that the current LEP phase of the CERN Large Electron-Positron
(LEP) collider represents the end of the road, as far as $e^+e^-$ storage
rings are concerned. To achieve higher energies with electrons, one has to
build a {\em linear} 
collider, but this naturally leads to a drop in luminosity,
as multiple bunch crossings are no longer possible.  Various designs for
the Next Linear Collider (NLC) are all subject to this disadvantage.  The
remedy -- to have $e^-$ and $e^+$ sources of extremely high intensity 
--- is obvious, but is expensive and subject to various technical 
limitations.

Muons represent an interesting alternative to electrons in a storage ring,
since they have similar quantum numbers but a mass about 200 times greater
than the electron mass --- thus  reducing synchrotron radiation a billion
times. Hence it should be possible to accelerate muons to much higher
energies in a storage ring. This is the prime motivation for building a
muon collider.  At such energies, the muon lifetime increases sufficiently 
to allow
several bunch crossings (typically a few hundred), leading to high design
luminosities.

The chief problem in machine design for a muon collider is the fact
that the muon, unlike the particles that have been used
till now, is an unstable particle, which, despite its increased lifetime at
high energies, ultimately tends to decay into an electron and a pair of
neutrinos. Thus, as a bunch of muons goes round and round the storage
ring, it gradually loses its muon content. Moreover, the electron
contamination tends to defocus the beam and introduce a momentum spread,
which calls for novel beam-cooling techniques (which are yet to be properly
devised). For energies around 300--500 GeV, which have been suggested for
the so-called First Muon Collider (FMC), muon decay tends to be a serious
limitation to the achievable luminosity. For a higher-energy machine of
2--4 TeV --- the Next Muon Collider (NMC) --- the muon lifetime increases
significantly and much larger luminosities should be achievable. However,
beam cooling becomes an even more serious problem here, and until more
concrete ideas about the technology become available, it is not clear that
a muon machine would be practical and cost-effective. There is also the
question of radiation hazards from neutrinos emitted in muon decay: this
crosses acceptable safety levels in the neighbourhood of the storage ring at
energies of the order of 3 TeV or more.

In view of the technical and financial constraints on building a muon
collider, it is important to make a serious study of the physics
capabilities of such a machine, especially when compared with a high
energy $e^+e^-$ machine such as the NLC. Only if a muon collider can
yield substantial improvements over the NLC in the physics output will
it be worth investing the time and money required to design and build it.
Thus it is important not only to study a muon collider from the point of
view of physics capabilities, but also from a comparative point of view
{\em vis-\`a-vis} $e^+e^-$ machines.

From the point of view of the physics probed, apart from Higgs-boson
Yukawa couplings where the greater mass of the muon plays an important
role, the other obvious place to look for differences between a machine
with muon beams and one with electron beams is in couplings that
differentiate between lepton flavours. Such couplings are absent from the SM
but may arise in supersymmetric models in three possible ways, {\em i.e.}
in the soft SUSY-breaking masses of sleptons, in trilinear slepton Higgs
boson couplings, and in $R$-parity-violating couplings, where lepton number
is violated. It is the last option that concerns us in this work.

\section{$\!\!\!\!\!\!.$~~$R$-Parity Violation}

Within the perturbative domain of the Standard Model, both baryon number
($B$) and lepton number ($L$) are conserved quantum numbers corresponding
to global $U(1)$ symmetries. Moreover, there is no leptonic-flavour
violation. This, however, is not true for the most general supersymmetric
Lagrangian consistent with the $SU(3)_c \times SU(2)_L \times U(1)_Y$
gauge symmetry of the SM.  The origin of this can be partly traced to the
fact that one of the doublet Higgs superfields ($H_1$) has the same gauge
quantum numbers as the doublet lepton superfields $L_i$ (containing
left-handed leptons) and may thus be replaced by any of the latter in the
superpotential. Moreover, trilinear terms involving the $SU(2)$-singlet
quark superfields (which incorporate right-handed quarks) are also
allowed. The additional pieces in the superpotential may thus be
parametrized~\cite{rpar}:
\be
    {\cal W}_{\not R} =   \mu_i L_i H_2 
                        + \l_{ijk}      L_i L_j \bar E_k
                + \l'_{ijk}     L_i Q_j \bar D_k
                + \l''_{ijk} \bar U_i \bar D_j \bar D_k,
      \label{R-parity}
\ee
where $\bar E_i, \bar U_i, \bar D_i$ are the singlet superfields and $L_i$
and $Q_i $ are the $SU(2)_L$-doublet lepton and quark superfields. Although
we have dropped $SU(2)_L$ and $SU(3)_c$ indices from the above expression,
it is important to note that the coefficients $\lambda''_{ijk}$ are
antisymmetric under the interchange of the last two flavour indices
because of the colour symmetry, while the $\lambda_{ijk}$ are
antisymmetric under the interchange of the first two flavour indices
because of $SU(2)_L$ invariance.

Of the four terms on the right-hand side of 
Eq.~(\ref{R-parity}), the first three\footnote{The
bilinear terms may be rotated away by a redefinition of the fields $(L_i,
H_1)$, with the effect reappearing as $\lambda$'s, $\lambda'$'s and,
possibly, lepton-number-violating soft supersymmetry-breaking
terms~\cite{bilinear}.} violate lepton number, while the last violates
baryon number. Simultaneous violation of $B$ and $L$ can, however, lead
to catastrophically high rates for proton decay.  A simple solution is the
introduction of a discrete symmetry with a conserved quantum number known
as $R$-parity. Expressible as $R_p \equiv (-1)^{3B + L + 2 S}$, where $S$
is the intrinsic spin of the particle~\cite{rpardef}, it cures the problem
of rapid proton decay~\cite{rpar} by forbidding {\it all} the terms in
Eq.~(\ref{R-parity}). A particularly interesting consequence of this
symmetry is that superpartners of SM particles always appear in pairs at
any vertex of the theory and hence the lightest supersymmetric particle
(LSP) must be absolutely stable. This has the advantage of providing a
possible dark-matter candidate. Moreover, the LSP is expected to interact
only weakly with matter, escaping most detectors and leading to rather
distinctive missing energy and momentum signatures at high-energy
colliders.

Although attractive, the requirement of $R$-parity invariance is evidently
an {\em ad hoc} one and has no compelling theoretical motivation.
Constraints from proton decay can be trivially evaded by demanding, for
example, that the theory respects $B$ but not necessarily $L$. In fact,
such an assumption is rather well-motivated within certain theoretical
frameworks~\cite{HalSuz, IbRo_92},  and for the remainder of our article
we shall consider $B$ to be a good quantum number of the theory, while $L$
is violated. In fact, vanishing of the $\lambda''_{ijk}$ couplings renders
preservation of GUT-scale baryogenesis~\cite{baryo} much easier than in
the case when the couplings are finite. Of course, the presence of the 
$L$-violating
terms can also affect the baryon asymmetry of the Universe, but the
consequent bounds are highly model-dependent and can easily be evaded, for
example by conserving just one lepton flavour over cosmological time
scales~\cite{DrRo_93}.

The last statement draws our attention to the issue of the simultaneous
presence of more than one $R$-parity-violating (\rp) coupling. This usually
results in tree-level flavour-changing neutral currents (FCNCs) and hence
there exist strong constraints on such scenarios~\cite{Products}, even though
{\em all} products of couplings are not constrained equally. Individual
couplings, on the other hand, can be constrained only from low-energy data
such as lepton or meson decays~\cite{BGH_89,BhCh_95}, neutrino
masses~\cite{GRT_93,BhCh_95}, neutrinoless double-beta
decay~\cite{DoubleBeta} and partial widths of the
$Z$-boson~\cite{z-decay}.  These constraints are neatly summarized in
Ref.~\cite{Dreiner}. As most of these constraints are relatively weak,
especially when the mass of the exchanged superparticle is large, they
leave enough room for remarkable signals at
LEP~\cite{GRT_93,BKP_95,lep2}, the Fermilab Tevatron~\cite{Rp_at_Tevat}
and future colliders. The recent observation of an excess in back-scattered 
positrons at the DESY HERA collider was initially thought \cite{HERA} 
to be just such a signal 
for $R$-parity violation, but it now appears more likely to have been a
mere statistical fluctuation \cite{Drees}.

In this article, we shall assume that one or at most two of the
$R$-parity-violating couplings are non-vanishing. This is certainly
making an assumption, but one can perhaps justify it by pointing at the SM
Yukawa couplings where the $t\bar tH$ coupling is completely
dominant. In any case, most of the effects discussed will not be changed
even if a third (or fourth) $R$-parity-violating coupling is non-zero,
though in such a case the low-energy bounds may be affected.

Under the given assumptions, then, $R$-parity violation can manifest
itself at a muon collider in one of three main ways:
\begin{itemize}
\item 
in pair-production of supersymmetric particles --- sfermions or gauginos
--- mainly through gauge interactions, followed by \rp\ decays;
\item 
in resonant production of a supersymmetric particle; and
\item 
in virtual effects in four-fermion processes.
\end{itemize}
Each of these possibilities requires a detailed analysis. The first
possibility would call for an analysis completely analogous to studies at
$e^+e^-$ colliders~\cite{GhGoRa} and will not be considered further in
this article. In any case, such a study would not be very useful in putting
bounds on the \rp\ couplings, since most of the results are very weakly 
dependent on the actual magnitude of these couplings.

To study the other two classes mentioned above, we must examine the
consequences of the terms in Eq.~(\ref{R-parity}). In terms of the
component fields, we have\footnote{We neglect effects due to quark
mixing. Such effects could lead to stricter constraints on the
couplings~\cite{Products}, but these can then be circumvented in various
ways, mostly by allowing some degree of fine tuning.}
\be
\barr{rcl}
    {\cal L}_{\lambda,\lambda'} & = & - \displaystyle
    \lambda_{ijk}  \left[{\tilde\nu}_{iL} \overline{\ell_{kR}} \ell_{jL} 
  + {\tilde \ell}_{jL} \overline{\ell_{kR}} \nu_{iL} 
  + {\tilde \ell}^\ast_{kR} \overline{(\nu_{iL})^c} \ell_{jL} 
  - (i \leftrightarrow j)\right] + {\rm H.c}
\\[2ex]
&   & \displaystyle
    -\lambda'_{ijk} \left[ \hspace*{.5em}
                       {\tilde \nu}_{iL} \overline{d_{kR}} d_{jL}
                     + {\tilde d}_{jL} \overline{d_{kR}} \nu_{iL}
                     + {\tilde d}^\ast_{kR} \overline{(\nu_{iL})^c} d_{jL}
               \right.
\\[2ex]
&  & \displaystyle \left. \hspace*{1.95em}
                     - \: {\tilde e}_{iL} \overline{d_{kR}} u_{jL}
                     - {\tilde u}_{jL} \overline{d_{kR}} e_{iL}
                     - {\tilde d}^\ast_{kR} \overline{(e_{iL})^c} u_{jL}
         \right] + {\rm H.c.}
\earr
\ee
We first concentrate on the possibility that only one \rp\ coupling is
non-zero, in which case it is easy to see that no tree-level FCNC effects
can be generated. In such a scenario, the most efficient probe is the
pair-production of charged fermions, {\em i.e.} processes of the type:
\be
    \mu^+ \mu^- \rightarrow f \bar{f} \ ,
    \label{pair-prodn}
\ee
where tagging of the final state is possible if $f = e,\mu,\tau,b$ and
(with rather low efficiency) $c$. It is important to be able to tag the
final states because this would enable us to know which of several
possible 
\rp\ couplings is involved. Tagging of leptons is rather straightforward,
with fairly high efficiencies possible even for taus; tagging of heavy
quarks ($b,c$) is more difficult since it involves observation of
displaced vertices; consequently the efficiencies are much lower. It is
not possible to tag light-quark flavours; in fact, if the final states
involve a pair of light quarks, an excess in dijet final states could
certainly be indicative of some new physics, but it would be difficult to
identify the particular $R$-parity-violating coupling(s) responsible. The
possibility of top-quark pair-production needs to be dealt with separately
and will not be discussed in this article.

\section{$\!\!\!\!\!\!.$~~Individual $R_p\!\!\!\!\!\!/$~~Couplings}

The $R$-parity-violating contribution to four-fermion processes in which 
the final-state fermions are not muons will come
from a $t$-channel exchange of a sneutrino or a squark, as the case may
be. For a $\mu^+ \mu^-$ final state, there will be an additional
$s$-channel
sneutrino exchange, with the possibility of a resonance (if the sneutrino
mass is in the accessible range). In Table 1, we have listed the \rp\
couplings that --- existing singly ---
can give rise to four-fermion processes
at a muon collider, together with the exchanged sfermion and the 
final-state particles. We also list there the current experimental
bounds~\cite{Dreiner} on these couplings for a light (heavy) sparticle
scenario. The $t \bar t$ final states, although they are not considered in
this article, are included in the table for the sake of completeness. The
upper half of the table is devoted to $\lambda$ couplings, while the lower
half deals with $\lambda'$ couplings. It is at once apparent that if the
single dominant coupling is any one of $\lambda_{131}, \lambda_{133}$ or
$\lambda'_{1jk}, \lambda'_{3jk}, (j,k = 1,2,3)$, then it will not
contribute to the four-fermion processes under consideration. These
couplings cannot, therefore, be measured in isolation
at a muon collider. However, the
other ones {\em will} contribute, and may be measured with varying degrees
of precision, depending on the final state. The precision will, of course,
depend on the efficiency of the tagging algorithm, for which a detailed
experimental simulation is required. In the absence of such studies, we
have assumed some typical efficiencies for flavour-tagging, which are guided
by LEP efficiencies, but are assumed to be 
uniform over the entire angular range.

The other crucial
factor affecting the measurement is the question of backgrounds.
It is, of course, immediately obvious that each of these processes will
have very large SM backgrounds. The SM contribution arises from 
$\gamma,Z$-mediated $s$-channel diagrams and, 
for $f = \mu$, additional $t$-channel
ones too (this process being the muonic analogue of Bhabha scattering).
Thus, a measurement of the production rate alone is unlikely to be the
most  sensitive test for an $R$-parity-violating contribution.  Of course,
in the fortuitous case when the
machine energy happens to hit on or near a resonance,
one can expect a large extra contribution to the cross section for 
$\mu^+ \mu^-
\rightarrow \mu^+ \mu^-$. This can, then, lead to rather good measurements of
the sneutrino mass and the responsible $R$-parity-violating 
coupling~\cite{Feng}. However, one must also allow for the possibility
that the sparticle resonance lies beyond the energy reach of the muon
collider and hence only virtual effects will be observable. These represent
smaller deviations from the SM cross sections, and, as such, require a more 
sensitive test to isolate from the background.

\vspace{-0.2in}
\footnotesize
\begin{center}
$$
\begin{array}{|c|c|c|c|c|}
\hline
{\rm Coupling}
& {\rm Final~state} & {\rm Exchange} & {\rm Channel(s)}
& {\rm Upper~bound}  \\
\hline
\lambda_{121} & e^+e^-       & \widetilde{\nu}_e    &   t & 0.05~(0.5) \\
\hline
\lambda_{122} & e^+e^-       & \widetilde{\nu}_\mu  &   t & 0.05~(0.5) \\
              & \mu^+\mu^-   & \widetilde{\nu}_e    & s+t &            \\
\hline
\lambda_{123} & \tau^+\tau^- & \widetilde{\nu}_e    &   t & 0.05~(0.5) \\
\hline
\lambda_{131} &    -         &          -           &  -  &            \\
\hline
\lambda_{132} & e^+e^-       & \widetilde{\nu}_\tau &   t & 0.06~(0.6) \\
              & \tau^+\tau^- & \widetilde{\nu}_e    &   t &            \\
\hline
\lambda_{133} &   -          &          -           &   - &            \\
\hline
\lambda_{231} & e^+e^-       & \widetilde{\nu}_\tau &   t & 0.06~(0.6) \\
\hline
\lambda_{232} & \mu^+\mu^-   & \widetilde{\nu}_\tau & s+t & 0.06~(0.6) \\
              & \tau^+\tau^- & \widetilde{\nu}_\mu  &   t &            \\
\hline
\lambda_{233} & \tau^+\tau^- & \widetilde{\nu}_\tau &   t & 0.06~(0.6) \\
\hline
\hline
\lambda'_{1jk}&      -       &         -            &  -  &            \\
\hline
\hline
\lambda'_{211}& d \bar d     & \widetilde{u}_L      &   t & 0.09~(0.9) \\
              & u \bar u     & \widetilde{d}_R      &   t &            \\
\hline
\lambda'_{212}& s \bar s     & \widetilde{u}_L      &   t  & 0.09~(0.9) \\
              & u \bar u     & \widetilde{s}_R      &   t  &            \\
\hline
\lambda'_{213}& b \bar b     & \widetilde{u}_L      &   t  & 0.09~(0.9) \\
              & u \bar u     & \widetilde{b}_R      &   t  &            \\
\hline
\lambda'_{221}& d \bar d     & \widetilde{c}_L      &   t  & 0.18~(1.8) \\
              & c \bar c     & \widetilde{d}_R      &   t  &            \\
\hline
\lambda'_{222}& s \bar s     & \widetilde{c}_L      &   t  & 0.18~(1.8) \\
              & c \bar c     & \widetilde{s}_R      &   t  &            \\
\hline
\lambda'_{223}& b \bar b     & \widetilde{c}_L      &   t  & 0.18~(1.8) \\
              & c \bar c     & \widetilde{b}_R      &   t  &            \\
\hline
\lambda'_{231}& d \bar d     & \widetilde{t}_L      &  t   & 0.22~(2.2) \\
              & t \bar t     & \widetilde{d}_R      &  t   &            \\
\hline
\lambda'_{232}& s \bar s     & \widetilde{t}_L      &  t   & 0.39~(3.5) \\
              & t \bar t     & \widetilde{s}_R      &  t   &            \\
\hline
\lambda'_{233}& b \bar b     & \widetilde{t}_L      &  t   & 0.39~(3.5) \\
              & t \bar t     & \widetilde{b}_R      &  t   &            \\
\hline
\hline
\lambda'_{3jk}&   -          &         -            &  -   &            \\
\hline
\end{array}
$$
\end{center}
\noindent
{\normalsize\rm Table 1}. {\footnotesize\em
List of $R$-parity-violating couplings and four-fermion processes to which
a single dominant coupling can contribute.  The exchanged sparticle is
shown, together with the current experimental bound.  The numbers shown
correspond to the case when the exchanged sparticle (relevant to the
bound) has a mass of 100 GeV (1 TeV) and the perturbative limit $\sqrt{4
\pi} \sim 3.5$ if there is no bound.  If the couplings do not lead to
four-fermion processes at a muon collider, the bounds (even where they
exist) are not listed.  }
\normalsize
 
Whatever the situation may be, and especially if the latter case should turn
out to be true, it is often more useful to consider the differential cross
sections for four-fermion processes rather than the overall rates. A
comparison of the experimentally measured angular distributions of leptons
or jets with those predicted in the SM might give a hint of the presence
of \rp\ contributions that are lost in the integrated cross sections. For
all the processes, except those leading to a $\mu^+ \mu^-$ final state,
the \rp\ contribution arises in the $t$-channel, whereas the SM background
arises from $s$-channel processes. Thus, one would expect a strong forward
peak in the \rp\ contribution, which is absent from the SM
background. For the $\mu^+ \mu^-$ final states, both signal and background
will have both $t$- and $s$-channel sfermion exchanges, so that the
difference in angular distribution may not prove so striking, but here the
$s$-channel resonance is likely to yield large excess contributions to the
new physics effect, even when the machine energy is not exactly tuned to 
the resonance.

In order to obtain a measure of the angular distribution, 
we integrate the differential
cross section over the scattering angle in bins of $5^\circ$
each for leptonic final states and $10^\circ$ each for hadronic final
states. Of course, in practice, hadronic jets would extend over several
bins, but we assume that the jet direction 
--- as determined by its thrust axis ---
can be determined to a precision of $\pm 5^\circ$. This assumption enables us
to make a parton-level Monte Carlo analysis of the problem. A more refined
analysis is certainly required at a later stage, but should not make
qualitative changes in the results shown here.

The number $N_i$ of events predicted in a bin $i$ (assuming a given \rp\
coupling and a sneutrino mass) is obtained by multiplying the integrated
cross section in that bin by the machine (integrated) luminosity $\cal L$
and the detection efficiency $\epsilon_{f \bar f}$ for the particular
fermion pair under consideration,
\be
    N_i = \epsilon_{f \bar f} {\cal L} \sigma_i \ .
 \label{eventno}
\ee
 
We then calculate the corresponding number $N_i^0$ in the SM for each bin
and use the two sets of numbers to construct the variance
\be \displaystyle
    \chi^2 = \sum_{i}^{ {\rm bins} }
      \left( \frac{ N_i - N_i^0}{\Delta N_i^0 } \right)^2\ .
  \label{chi2}
\ee
The error $\Delta N_i^0$ in Eq.~(\ref{chi2}) is obtained by adding the
statistical error and a projected systematic error $\delta N_i^0$ in
quadrature:
\be
\Delta N_i^0 = \sqrt{ (\sqrt{N_i^0})^2 + (\delta N_i^0)^2 }.
  \label{error}
\ee
This systematic error is a theoretician's apology for an experimental
analysis and we take it to be 2\% for all the bins, {\em i.e.}
$\frac{\delta N_i^0}{N_i^0} = 0.02$. Of course, the actual figure may be
somewhat different from this and will probably vary over different bins.
However, we include it as a zeroth-order estimate. Obviously $\chi^2$
grows as the mismatch between the two distributions grows and can be taken
as a measure of the new-physics effect.

To be quantitative, we adopt the energy and luminosity parameters given by
the Accelerator Physics Study Group at the Fermilab Workshop on the First
Muon Collider~\cite{FNAL}. For the FMC, these are
\begin{eqnarray}
\sqrt{s} &=& 350,~500 \gev; \nonumber \\
{\cal L} &=& 10 \fb^{-1},
\end{eqnarray}
which we shall denote FMC-350 and FMC-500 respectively. For the NMC these
parameters are
\begin{eqnarray}
\sqrt{s} &=& 2,~4 \tev; \nonumber \\
{\cal L} &=& 10^3 \fb^{-1}, 
\end{eqnarray}
which we shall denote NMC-2 and NMC-4 respectively.

Accurate flavour tagging demands that the outgoing leptons or jets in
Eq.~(\ref{pair-prodn}) be at least $20^\circ$ away from the beam pipe.
This is indicated by preliminary studies \cite{FNAL}
of the electron contamination in
the initial muon beam. Within this restricted region, we now assume
uniform detection efficiencies~\cite{tariq}
\be
\epsilon_{ee} = \epsilon_{\mu\mu} = 0.9, \qquad 
\epsilon_{\tau\tau} = 0.8,
\qquad
\epsilon_{b \bar b} = 0.3, \qquad \epsilon_{c \bar c} = 0.1.
     \label{efficiencies}
\ee
These are again {\em ad hoc} assumptions (albeit guided by LEP
efficiencies), but they cannot really 
be improved until the detector is actually
designed. 

Dividing the angular region between $20^\circ$ and $160^\circ$
into equal-sized bins of $5^\circ(10^\circ)$ each for leptonic (hadronic)
final states leads to 28 (14) bins. To avoid spurious contributions to the
variance $\chi^2$ we drop a bin from the sum in Eq.~(\ref{chi2}) if either
$(i)$ the difference between the SM expectation and the measured number of
events is less than 1 or $(ii)$~the SM expectation is less than 1 event.
However, the high luminosities ensure that this does not affect the numerical
results significantly. We then 
present the discovery limits from such an analysis in the form of 95\%
C.L. contours in the space of the sfermion mass and the \rp\ coupling. A
signal will be deemed observable if the $\chi^2$ of Eq.~(\ref{chi2})
exceeds the minimum
acceptable fluctuation~\cite{pdg} for 28 (14) random variables. Before we
discuss our results, a few observations are in order.
\begin{itemize}
\item Throughout our analysis, we shall assume that only one sfermion (and
its isospin partner, if any) is `light'. In other words, that the
left-chiral and right-chiral sfermions have widely different masses.
Hence we shall not combine limits from processes with differing sfermion
propagators.  
\item Although it might seem inappropriate to consider four-fermion
processes mediated by sfermions that are light enough to be pair-produced,
it is not quite so. For, even in the case of a pair of sfermions being
produced, a determination of the strength of possible \rp\ interactions
from a study of the branching fractions is likely to be very difficult.
The four-fermion mode can then play a complementary role in pinning
down the actual value of the \rp\ coupling.
\item All our results for dijet final states are based on a parton-level
analysis. While jet fragmentation details may affect the results somewhat,
the qualitative features will remain unchanged.  
\item We neglect all initial-state radiation (ISR) effects. This, of
course, is a much better approximation for a muon collider than for an $e^+
e^-$ one. Some small changes may occur, though probably not enough to
change the qualitative features of our work. However, if it is proposed to
tune a muon collider to a sneutrino resonance (assuming it has previously
been discovered), then ISR effects could be critical.
\item In an actual experimental study, one must demand that the energies
of the final lepton or jet pair match the beam energy well. This would
serve to remove both ISR effects and backgrounds from SM (or MSSM)
processes where one or more particles escape undetected.
\end{itemize}

Once all these caveats and approximations 
are taken into account, Fig.~1 shows
the discovery limits for the $\l$ couplings, obtainable by studying
dilepton states, while Figs.~2 and 3 show the discovery limits for the
$\lp$ couplings, obtainable by studying dijet states. In view of the
large number of curves in each figure, and the different possibilities for 
\rp\ couplings, these cases merit separate discussion.

\noindent
\underline{\it Dilepton final states:}

In Fig.~1($a$), we illustrate the limits achievable from $\mu^+ \mu^-
\ra \mu^+ \mu^-$ for the relevant couplings (see Table 1) $\l_{122}$ and
$\l_{232}$ with $\tilde \nu_e$ and $\tilde \nu_\tau$, respectively, 
being exchanged in both $s$ and $t$ channels. The
current low-energy bounds on these are shown by (solid) parallel straight
lines in the figure, the upper one denoting $\l_{232}$ and the lower one
denoting $\l_{122}$. The dashed line shows the reach in the
$\l$--$M_{\widetilde{\nu}}$ plane for the coupling $\l_{122}$ at LEP
($\l_{232}$ cannot be measured in isolation 
at an $e^+ e^-$ collider) running at a
centre-of-mass energy of 190 GeV and accumulating 300 pb$^{-1}$ of data,
the last being a fairly conservative assumption for the LEP collider when
the data of all four experiments are combined. Of course, for the LEP
bounds on the $\l_{122}$ coupling, one needs to study\footnote{See also
Section 6, where the case of $e^+e^-$ colliders is discussed at length.}
the $t$-channel sneutrino exchange process $e^+e^- \ra \mu^+\mu^-$.  
Solid lines show the discovey reach at
the FMC-350 and FMC-500, and at the NMC-2 and NMC-4.

\begin{figure}[h]
\begin{center}
\vspace*{4.2in}
      \relax\noindent\hskip -4.2in\relax{\includegraphics{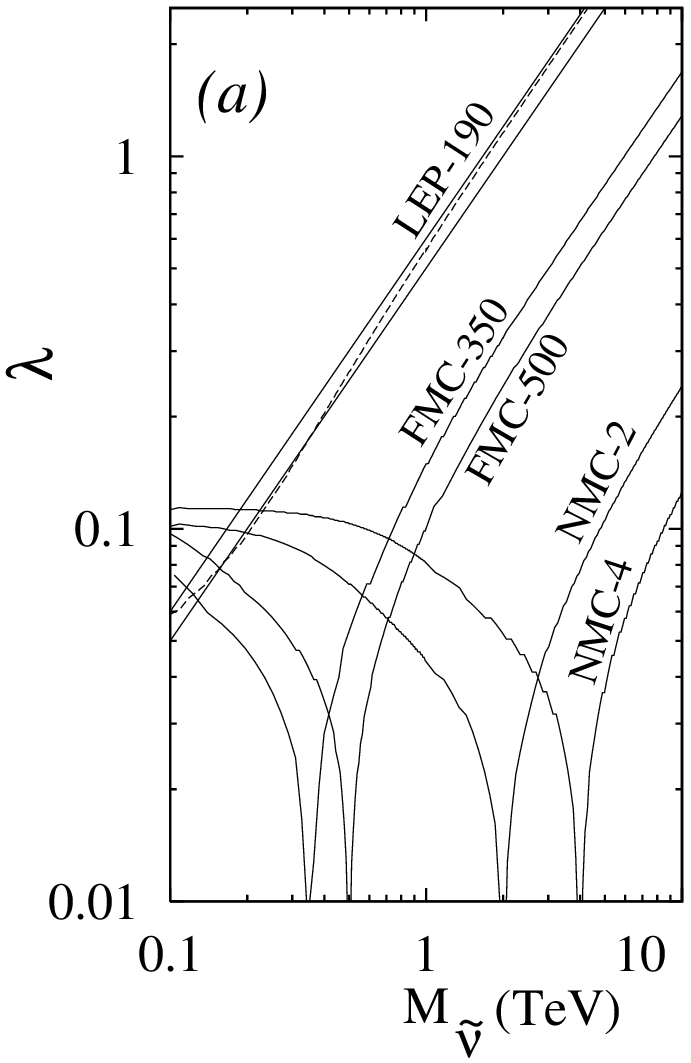}}
                     \hskip 1.99in\relax{\includegraphics{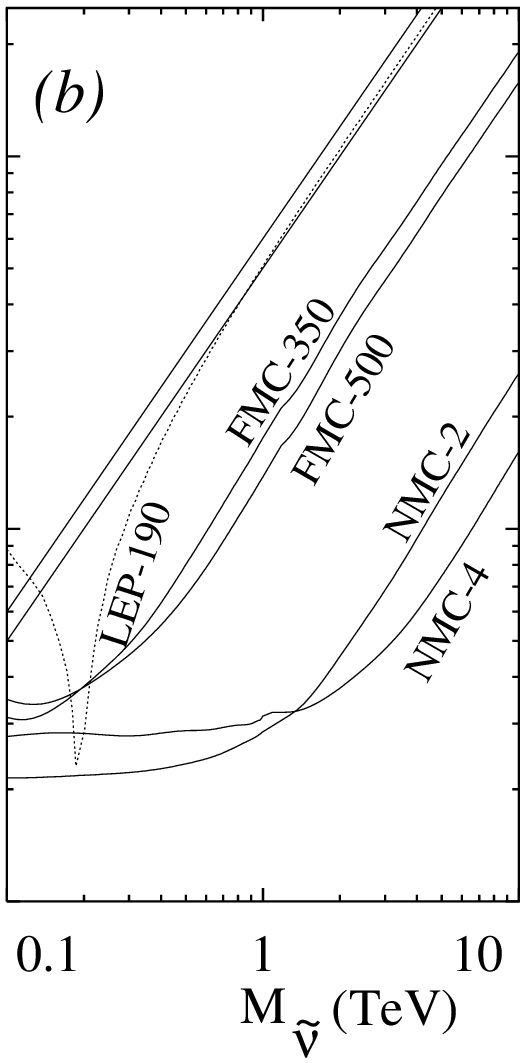}}
                     \hskip 2.005in\relax{\includegraphics{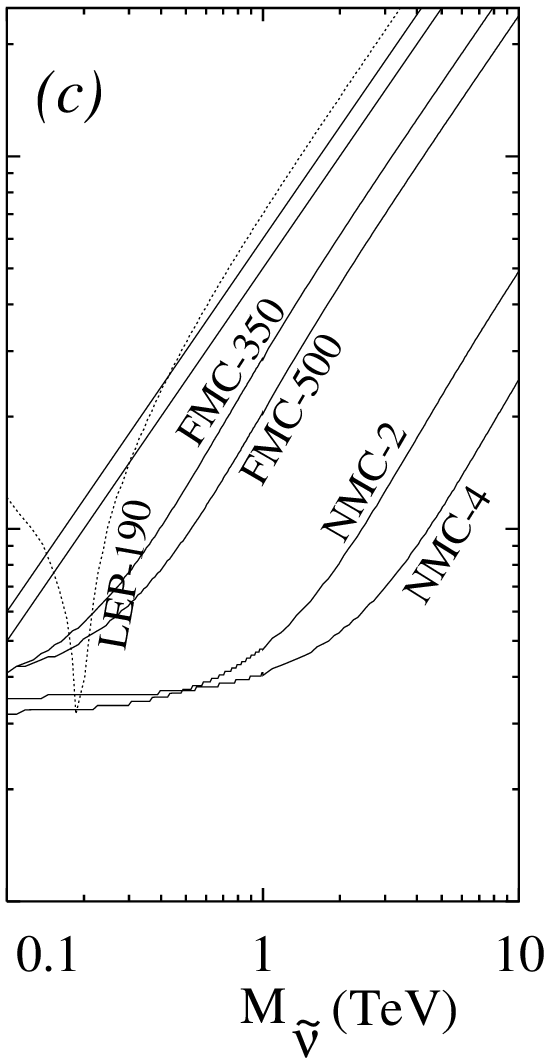}}
\end{center}
\vspace*{-1pt}
\caption{\footnotesize\em
Illustrating the reach at a muon collider in the \rp\ parameter space
($LL\bar E$) when ($a$) the differential cross section for a $\mu^+\mu^-$
final state is considered, ($b$) the differential cross section for an
$e^+e^-$ final state is considered, and ($c$) the total cross sections for
an $e^+e^-$ final state are considered. The part of the parameter space
{\em above} the curves can be excluded at 95\% C.L. The oblique straight
lines correspond to low-energy bounds on the relevant couplings (see
text). }
\end{figure}

The sharp dips in the contours correspond to the resonance production of a
sneutrino that subsequently decays into $\mu^+ \mu^-$ through the same
coupling. As mentioned above, 
the large cross section at the resonance enables us to probe
much smaller values of the coupling than when the sneutrino is an
off-shell one. For the purposes of this graph, the sneutrino width is taken to
be $200$ MeV. Such a value assumes that the sneutrino decays almost
exclusively through \rp\ channels, which may not always be
true, since it is possible for the sneutrino to decay to
lepton--gaugino final states. This assumption automatically restricts us
to the part of the MSSM parameter space where gauginos are heavier than
the sneutrino. A complementary study by Feng, Gunion and Han~\cite{Feng}
discusses the situation when the gauginos are light.

Fig.~1($a$) makes it clear that it should be possible, at a muon collider,
to place rather stringent bounds\footnote{In this and the subsequent
discussions, we refer to an improvement of bounds, but it is, of course,
possible that a discovery may be made.} on the couplings $\l_{122}$ and 
$\l_{232}$, which go far beyond what can be inferred from low-energy 
data or from the data expected in the current run of LEP. In fact,
it is obvious that LEP can barely improve upon the low-energy bound on
$\l_{122}$, even when the sneutrino mass is 1 TeV or more. The situation
is clearly more promising at a muon collider, especially when the sneutrino
mass exceeds 200 GeV.  Much of this is due to resonance effects. In fact,
near the peak, bounds on $\l_{122}$ and $\l_{232}$ can be improved by a
whole order of magnitude. Even away from the resonance, if a muon
collider is built and runs in all the above modes, one can rule out values
of $\l > 0.06$ all the way up to a sneutrino mass of 5--6 TeV.

The situation is somewhat different when a resonance is
absent, as Fig.~1($b$) and 1($c$) demonstrate. Fig.~1($b$) shows the
discovery limits for an $e^+ e^-$ final state, which implies one of the
couplings $\l_{121}$, $\l_{122}$, $\l_{132}$ or $\l_{231}$. Once
again, the oblique straight lines demarcate the region ruled out by
low-energy data; the lower line is relevant for $\l_{121},\l_{122}$
and the upper one for $\l_{132},\l_{231}$. The dashed line shows
the bounds available from LEP on $\l_{121}$, which can cause a sneutrino
resonance in Bhabha scattering at LEP (see Section 6). 
We see that muon colliders could again do
better than the low-energy data, especially in the higher sneutrino mass
region.  For the $\l_{122}$ coupling, a combination of Fig.~1($a$) and
($b$) can reduce the bound to about 0.015 for sneutrino masses upto a
few TeV. It is interesting that one gets, in general, better bounds from the
non-resonant case of $e^+ e^-$ final states than can be obtained from the
case of $\mu^+ \mu^-$ final states (unless, indeed, the machine energy
happens to hit on a sneutrino resonance in the latter case). This is
because of the extra $t$-channel diagrams in the SM background for the
$\mu^+ \mu^-$ final state, which make the signal and background look
rather similar, as far as angular distributions go.

To illustrate the usefulness of comparing differential cross sections, in
Fig.~1($c$) we have shown the discovery limits obtainable from a
consideration of the total Bhabha 
cross section for $\mu^+ \mu^- \rightarrow e^+
e^-$. We simply calculate the fluctuation in the total SM cross section by
a formula identical to Eq.~(\ref{error}) (assuming a 2\%
systematic error) and require the excess supersymmetric contribution to
exceed this at 95\% C.L. As before, dashed lines show the bounds available
from LEP, from a consideration of the total cross section only. It is
immediately obvious that comparison of differential cross sections by the
method of Fig.~1($b$) leads to better discovery limits than a
consideration of the total cross section. In fact, with the latter
approach, the LEP bounds would hardly better the low-energy bounds (unless
a resonance is hit upon), and though the FMC could improve these, it will
cover less of the parameter space than that shown in Fig.~1($b$). The
NMC, of course, does pretty well in either case,
simply by virtue of its high energy and
luminosity, but again, this method fares somewhat worse than that of
Fig.~1($b$).

Both Figs.~1($a$) and 1($b$) show a crossover between the two lines
representing the FMC discovery limits. A similar phenomenon occurs for the
NMC. This implies that, for lower values of sneutrino mass, it is
actually more efficient to have a lower-energy machine than a higher-energy 
one (with the same luminosity). 
This apparent paradox is actually an artefact of the cuts. For higher
energies, the sneutrino mass becomes progressively negligible, and thus the
excess contribution, being driven by a $t$-channel exchange, is more and
more peaked at small angles. Since precisely this part of the distribution
is rendered inaccessible by detector limitations, the relative excess
is reduced at higher energies.

An analysis of the $\tau^+ \tau^-$ final state results in very similar
plots for the couplings $\l_{123}$, $\l_{132}$, $\l_{232}$ or $\l_{233}$
(the last two are inaccessible at LEP). The small differences reflect
the fact that, at a muon collider, tau-tagging efficiency is expected
to be close to that for tagging muons (SEe Eq.~(\ref{efficiencies})).  We,
therefore, choose not to present these plots.  It must be remembered,
however, that for the $\l_{123}$ and $\l_{132}$ couplings at LEP, 
a $t$-channel sneutrino
exchange is relevant and hence we need to replace the dotted curve of
Fig.~1($b$) with the dashed curve of Fig.~1($a$).

\noindent
\underline{\it Dijet final states:}

We now turn to the possibility of dijet final states, where effects due to
$\lp$ couplings could manifest themselves. These are qualitatively similar
to the $\mu^+ \mu^- \rightarrow e^+ e^-$ case, since the new-physics
effect arises from the replacement of a $t$-channel sneutrino by a
$t$-channel squark. There are quantitative differences, however. One is
the obvious fact that the cross sections are changed because of the colour
factor and the different weak quantum numbers of quarks. A greater
difference arises because tagging efficiencies for heavy $b,c$ quarks are
low and the lighter flavours cannot be tagged at all. Without tagging, we
can tell very little about which individual coupling is responsible for
the excess events, if they are seen. Of course, if no deviation is seen,
{\em all} the couplings that could have resulted in such an excess are
automatically bounded, but the converse is certainly not true.

Perhaps it would not be out of place to explain how the tagging efficiency
affects the $\chi^2$ fitting of the distribution. It is apparent from
Eqs.~(\ref{eventno}) and (\ref{chi2}) that decreased efficiency decreases
the effective luminosity and thus increases the relative statistical error
in Eq.~(\ref{chi2}). As a result, it becomes more difficult to
differentiate the signal from the background through a $\chi^2$ analysis.
Another issue of great importance in the study of dijets is the question
of charge identification. Using the wisdom of LEP estimates and practice,
we assume that charge identification is possible for heavy quark jets and
that it 
may or may not be possible for light-quark jets. If charge identification
is possible, then the scattering angle $\theta$ is uniquely defined and
may be treated in the manner described above, dividing the range $20^\circ
< \theta < 160^\circ$ into 14 bins of $10^\circ$ each.  However, if charge
identification is not possible, then it will not be possible to
distinguish between $\theta$ and $\pi-\theta$; hence one should add to
the cross section in the bin $20^\circ < \theta < 30^\circ$ the
cross section in the bin $150^\circ < \theta < 160^\circ$, and so on.
Thus, we get a total of 7 bins, with a corresponding dilution of the
capacity to distinguish between signal and background. However, the
minimum $\chi^2$ requirement is also reduced, so there is a partial
trade-off between the two effects, as we shall see presently.

\begin{figure}[h]
\begin{center}
\vspace*{4.2in}
      \relax\noindent\hskip -4.2in\relax{\includegraphics{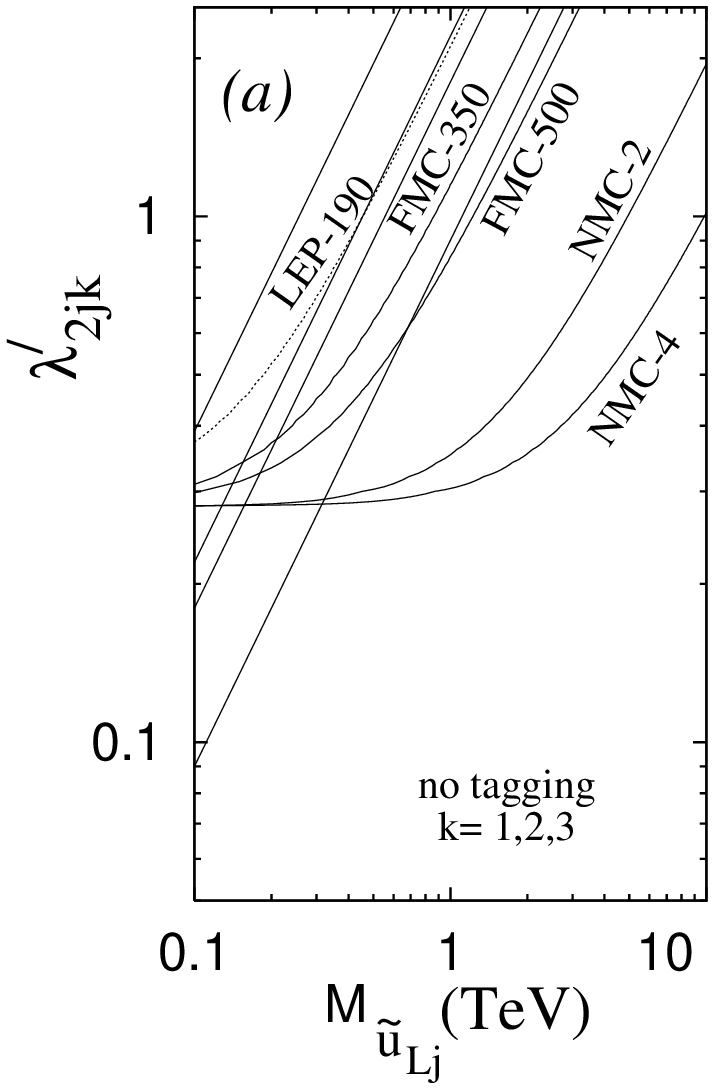}}
                     \hskip 1.99in\relax{\includegraphics{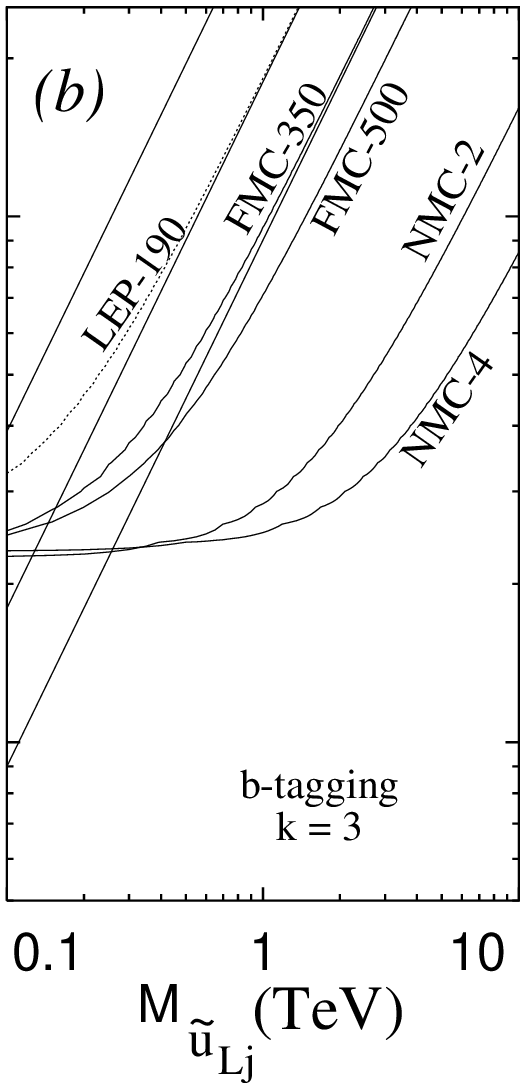}}
                     \hskip 2.005in\relax{\includegraphics{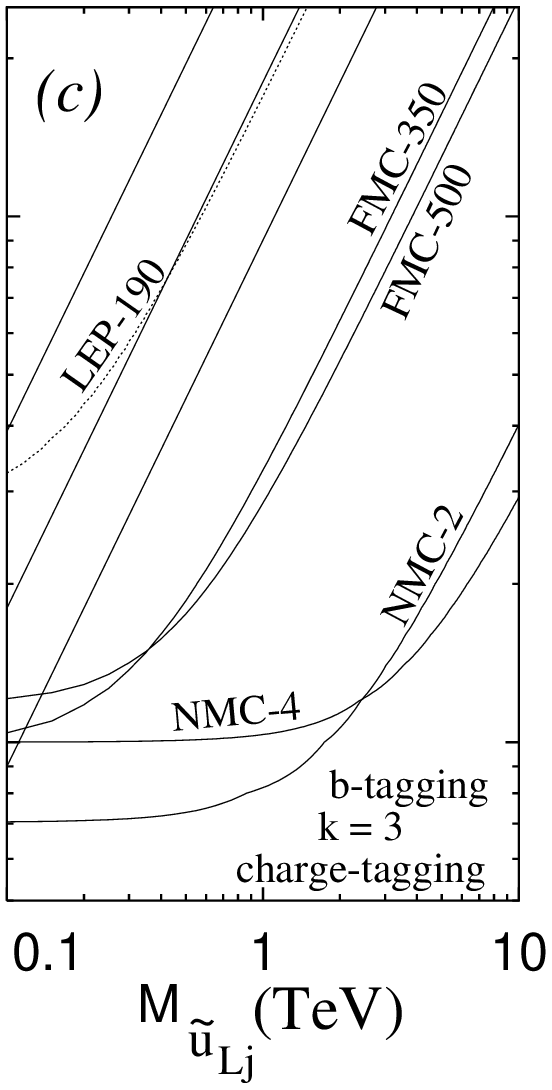}}
\end{center}
\vspace*{-1pt}
\caption{\footnotesize\em
Illustrating the reach of a muon collider in the \rp\ parameter space
($LQ\bar D$), assuming exchange of a $\tilde u_L$-type squark, when ($a$)
the differential cross section for an untagged dijet final state is
considered, ($b$) the differential cross section for a tagged $b \bar b$
final state is considered, but we assume no charge identification, and
($c$) the differential cross section for a tagged $b \bar b$ final state
is considered, assuming charge identification with unit efficiency. The
oblique straight lines correspond to low-energy bounds on the relevant
couplings (see text). }
\end{figure}

In Fig.~2($a$), we show the discovery limits obtainable for the
$\lp_{2jk}$ ($j,k=1,2,3$) couplings from a consideration of dijet final
states, where no attempt is made to tag the jets. For this figure, we
assume that the exchanged squark has charge $\frac{2}{3}$ and hence the
excess will appear in $d \bar d, s \bar s$ or $b \bar b$ states.  Of
course, {\em all} 
the light quarks, irrespective of charge, will form part of
the SM background. Naturally, there is no resonance involved here and
hence the curves look rather like those in Fig.~1($b$).  The difference
lies in the different coupling, the fact that we use a broader binning for
hadronic jets and the fact that there is no charge
identification. Most important, however, is the sheer magnitude of the SM
background, to which all light quarks contribute. All of
the $\lp$ couplings listed in Table 1 can contribute to this figure, and
hence the bounds from this are the most general of all. The dotted line
shows possible 
LEP bounds, but it should be remembered that these correspond
to a different set of 
couplings $\lp_{1jk}$. As in Fig.~1, the four oblique (solid) lines
show the low-energy bounds on the $\lp_{2jk}$ couplings, which may be
read off from Table 1. We have not shown the bounds on the $\lp_{1jk}$,
which are rather more tight than those on the $\lp_{2jk}$
(so that the figure is slightly misleading as regards the LEP bounds), 
but these may be readily found in Ref.~\cite{Dreiner}. Thus the figure
tells us that a muon collider can probe $\lp_{2jk} \gsim 0.25$ for
$M_{\snu} \lsim 1$ TeV.

In Fig.~2($b$), we show the discovery limits obtainable from a
consideration of $b \bar b$ final states, where both the $b$-jets are
tagged, but charge identification is not done. The bounds are valid for
the couplings $\lp_{2j3},\ (j = 1,2,3)$, as a glance at Table 1 will show.
The three straight lines in the top left-hand corner denote (in ascending
order), the bounds on $\lp_{2j3}$, where $j = 1, 2, 3$ respectively, and
the dotted line shows what LEP can achieve for the couplings
$\lp_{1j3}$, given 300 pb$^{-1}$ of data.  Again, the graphs resemble
those of Fig.~1($b$), for reasons explained above; but it is clear that
the low tagging efficiency for $b$-jets compared with that for electrons or
positrons leads
to less striking results at the FMC. At the NMC, the high luminosity comes
to the rescue, since the smallest difference in cross section now gets
magnified in the construction of $\chi^2$.  
It is also interesting that even with the rather low ($\sim
0.3$) efficiency of tagging $b$-jet pairs, there is a significant
difference between the accessible region with and without $b$-tagging. 
This may be attributed to a large decrease in the background with the
removal of other quark flavours leading to dijets.  A
much more striking improvement occurs once charge-identification is
assumed. This is illustrated in Fig.~2($c$), where it is further
assumed that
the charge of the tagged $b$-jets can be identified with 100\% efficiency.
Why is this s.~ To understand this feature, we must examine the angular
distribution of the $b$-jet, which is peaked in the forward hemisphere. On
the other hand, the {\em relative} deviation due to a small \rp coupling
is more pronounced in the backward hemisphere, although the cross sections
are considerably smaller. In the absence of charge identification, since
$\theta$ is indistinguishable from
$\pi - \theta$, we need to sum over the paired bins and
the relative deviation is thereby reduced\footnote{A numerical example will 
help clarify this. For a certain choice of parameters, the SM prediction
for the bin $40^\circ$--$50^\circ$ is 80 events, with 85 events for the
\rp\ case. The same choice leads to 12 SM events in the bin 
$130^\circ$--$140^\circ$, with 16 events for the \rp\ case. 
We thus get $\chi^2
= 0.303 + 1.327 = 1.63$ for the case when bioth bins are considered,
and $\chi^2 = 0.849$ when they are merged into a single bin.}. 
This, in turn, reduces the
contribution to the $\chi^2$, and thus the sensitivity to $\lp$.

In Fig.~3($a$), we again show the discovery limits obtainable on the
couplings $\lp_{2jk}$, ($j,k=1,2,3)$ from a consideration of dijet final
states, where no attempt is made to tag the jets. For this figure, we
assume that the exchanged squark has charge $-\frac{1}{3}$ and hence the
excess will appear in $u \bar u$ or $c \bar c$ states.  As before, all the
light quarks, irrespective of charge, will form part of the SM background.
The graph differs from Fig.~2($a$) only in that the
interference terms between the \rp-contribution and the SM ones lead to a
larger excess 
cross section because of the different quantum number assignments
of the final states. For this reason, the bounds could be significantly
better. In fact, we could then probe $\lp_{2jk} \gsim 0.1$ all the way
up to $M_{\snu} \simeq 1$ TeV.

\begin{figure}[h]
\begin{center}
\vspace*{4.2in}
      \relax\noindent\hskip -4.2in\relax{\includegraphics{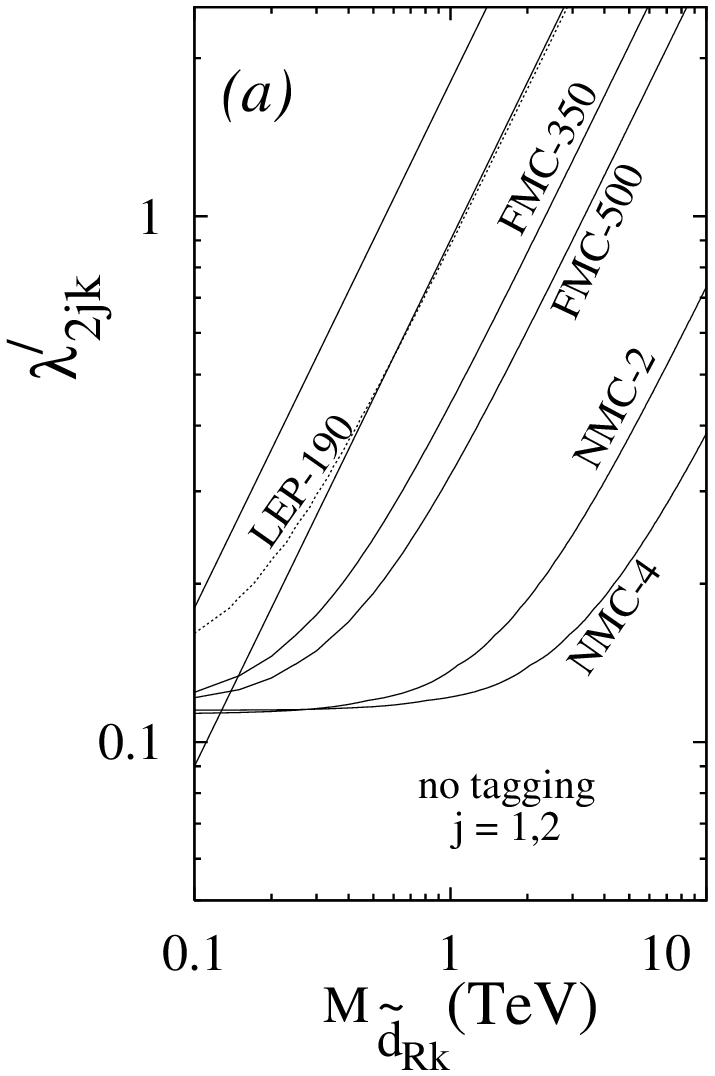}}
                     \hskip 1.99in\relax{\includegraphics{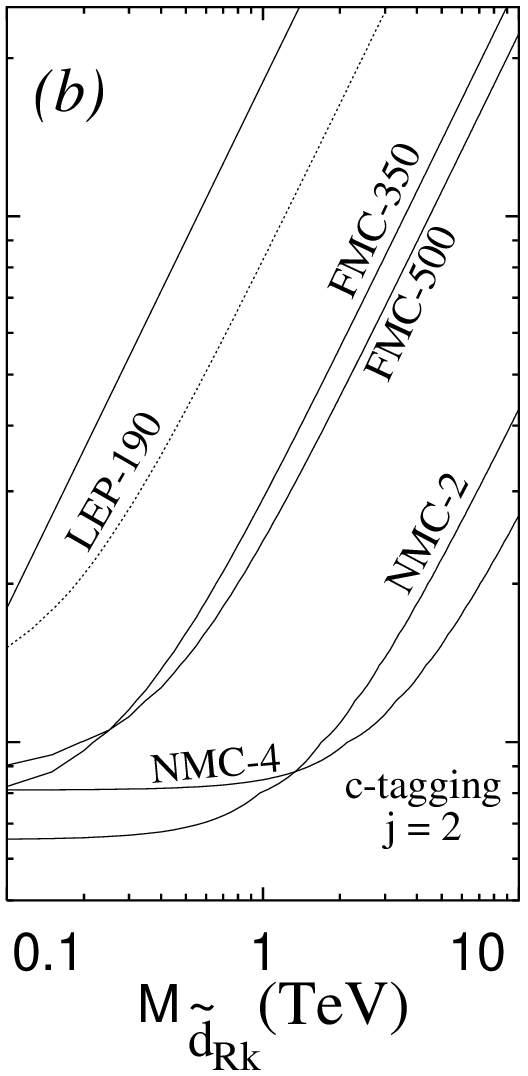}}
                     \hskip 2.005in\relax{\includegraphics{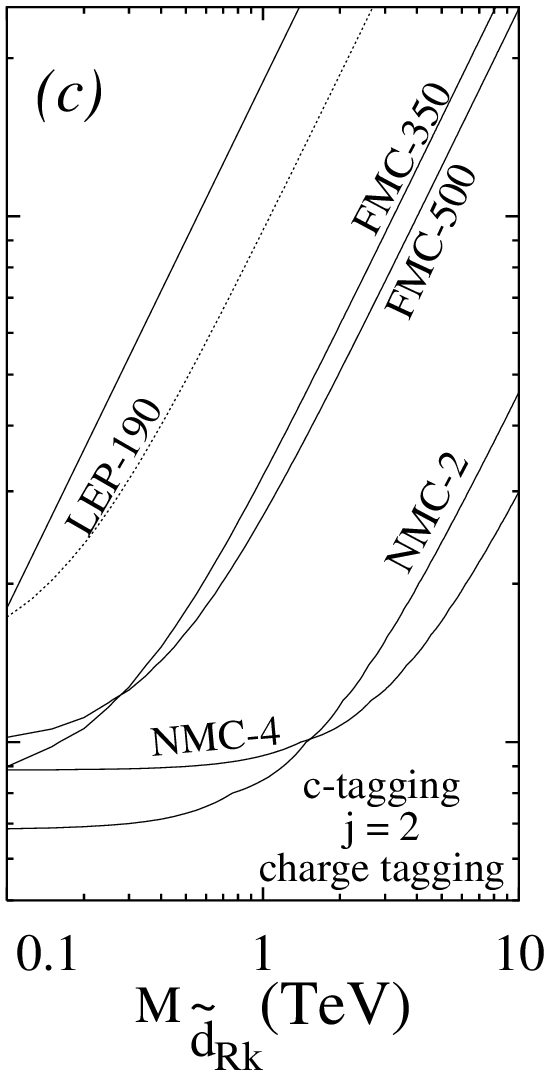}}
\end{center}
\vspace*{-1pt}
\caption{\footnotesize\em
As in Fig.~2, but assuming exchange of a $\tilde d$-type squark, and
tagged $c \bar c$ final states. }
\end{figure}

In Fig.~3($b$), as in Fig.~2($b$), we show the discovery limits
obtainable from a consideration of $c \bar c$ final states, where both 
jets are tagged as arising from $c$-quarks, but charge identification is
not attempted. This means the couplings involved must be $\lp_{22k}, (k =
1,2,3)$ (see Table 1).  Here the bounds are actually stronger than in the
case of $b$-jets, despite the efficiency for tagging $c$-quarks being
rather poor ($\sim 0.1$).  This is because the excess cross section for
charge $\frac{2}{3}$ final states is larger than that for charge
$-\frac{1}{3}$ ones, essentially on account of large interference
terms that are 
strongly dependent on the quantum-number assignments of the 
final-state 
quarks. Rather surprisingly, given the arguments presented above for
the case of $b \bar b$ final states, not much improvement is achieved by
identifying the charge of the parent $c$-quarks (Fig.~3($c$)). This
paradox is simply resolved by noting that, for the case of $u$-type quarks,
the difference between signal and background is more pronounced in the
forward hemisphere, where the cross section is larger. Thus the
major contribution to the $\chi^2$ comes from the forward hemisphere and
separating out the backward hemisphere has little effect on the result.
What little improvement is attained is, in fact, washed out by the
requirement of a larger $\chi^2$ when we double the number of bins.
Thus, it might be useful to simply look at the seven bins in
the forward hemisphere.
For $c \bar c$ final states, then, unlike the $b \bar b$ case, charge
identification is not a major issue.

We conclude this section with the remark that a comparison of Fig.~1
with Figs.~2 and 3 shows that high tagging efficiency does not
necessarily imply a better reach in the space of $R$-parity-violating
couplings. This reach 
depends on several features, and the quantum numbers of
the final-state fermions play a crucial role in its determination.

\section{$\!\!\!\!\!\!.$~~Products of $R_p\!\!\!\!\!\!/$~~Couplings}

We next consider the possibility that more than one of the couplings in
Eq.~(\ref{R-parity}) is non-zero. This situation is rather more
complicated. Clearly, one interesting possibility is to have tree-level
FCNCs, which would then be susceptible to various bounds from low-energy
processes. Many of these have been studied and various bounds may be found
in the literature~\cite{Products}.  In a muon collider, FCNC interactions
will manifest themselves in several ways, the simplest being in processes
such as $\mu^+ \mu^- \rightarrow f_i \bar f_j~(i \neq j)$, where $i$ and
$j$ label different flavours.  Experimentally, such mixed-flavour final
states are easiest to detect when $f_i,f_j$ are charged leptons. In this
case, the final state will be indicative of a product of two couplings
both of the $\lambda$ type, with an exchanged sneutrino. There will be
virtually no SM background for these processes and very small backgrounds
(if at all) from other supersymmetric processes involving lepton
flavour-mixing (since these occur at the one-loop level)~\cite{Cheng}.
Thus, one can simply search for unlike-flavour lepton pairs as a signal
for $R$-parity-violating scenarios where more than one coupling is
non-zero.

The recent results from Super-Kamiokande~\cite{Kajita}, indicating the
occurrence of neutrino oscillations, may perhaps be thought to lead to
mixing of charged lepton states as well, within the framework of the SM
itself.
However, even if such mixings occur, the effects must be so small as to
have evaded detection (to date) in all low-energy experiments.
The size of the detector is, of course,
much too small to see oscillations between charged leptons. Thus, such
effects would be negligible at a muon collider and would not interfere
with any signals due to \rp-interactions.

\vspace{-0.2in}
\footnotesize
\begin{center}
$$
\begin{array}{|c|c|c|c|c|}
\hline
{\rm Product} \: \Lambda^2& {\rm Final~state}& {\rm Exchange}
& {\rm Channel(s)} & {\rm Upper Bound} \\
\hline
\l_{121}\l_{122}& e\mu   &\widetilde{\nu}_e   &s+t& 6.6\times 10^{-7(5)}\\
\hline
\l_{121}\l_{123}& e\tau  &\widetilde{\nu}_e   &  t& 5.6\times 10^{-3(1)}\\
\hline
\l_{122}\l_{123}&\mu\tau &\widetilde{\nu}_e   &s+t& 5.7\times 10^{-3(1)}\\
\hline
\l_{122}\l_{131}&e\tau   &\widetilde{\nu}_e   &  s& 6.7\times 10^{-3(1)}\\
\hline
\l_{122}\l_{132}&\mu\tau &\widetilde{\nu}_e   &s+t& 6.4\times 10^{-3(1)}\\
\hline
\l_{122}\l_{133}&\tau\tau&\widetilde{\nu}_e   &  s& -                   \\
\hline
\l_{131}\l_{232}&ee      &\widetilde{\nu}_\tau&  s& -                   \\
\hline
\l_{132}\l_{232}&e\mu    &\widetilde{\nu}_\tau&s+t& -                   \\
\hline
\l_{133}\l_{232}&e\tau   &\widetilde{\nu}_\tau&  s& 5.6\times 10^{-3(1)}\\
\hline
\l_{231}\l_{232}&e\mu    &\widetilde{\nu}_\tau&s+t& -                   \\
\hline
\l_{231}\l_{233}&e\tau   &\widetilde{\nu}_\tau&  t& 6.7\times 10^{-3(1)}\\
\hline
\l_{232}\l_{233}&\mu\tau &\widetilde{\nu}_\tau&s+t& 6.4\times 10^{-3(1)}\\
\hline
\hline
\l_{i22} \lp_{i11},
\l_{i22} \lp_{i22} & {\rm dijet}  & \widetilde{\nu}_i & s & -  \\
\hline
\l_{i22} \lp_{i21},
\l_{i22} \lp_{i12} & {\rm dijet}  & \widetilde{\nu}_i & s &
         3.8 \times 10^{-7(5)}  \\
\hline
\l_{i22} \lp_{i13},
\l_{i22} \lp_{i23} & {\rm dijet~(one}~b) & \widetilde{\nu}_i & s & - \\
\hline
\l_{i22} \lp_{i33} & b \bar b & \widetilde{\nu}_i & s & - \\
\hline
\hline
\lp_{2i1} \lp_{2i2} & {\rm dijet}& \widetilde{u}_{iL} & t &
         3.8 \times 10^{-7(5)}  \\
\hline
\lp_{2i1} \lp_{2i3},
\lp_{2i2} \lp_{2i3} & {\rm dijet~(one}~b) & \widetilde{u}_{iL} & t & - \\
\hline
\lp_{21i} \lp_{22i} & {\rm dijet~(one}~c) & \widetilde{d}_{iR} & t & -  \\
\hline
\lp_{21i} \lp_{23i} & \leq 4~{\rm jets~(one}~b) & \widetilde{d}_{iR} & t & - \\
\hline
\lp_{22i} \lp_{23i} & \leq 4~{\rm jets}& \widetilde{d}_{iR} & t & - \\
                    & ({\rm one}~b,~{\rm one}~c) &                  &  &   \\
\hline
\end{array}
$$
\end{center}
\noindent
{\normalsize\rm Table 2}. {\footnotesize\em
List of products of $R$-parity-violating couplings that can be measured
in four-fermion processes at a muon collider. The exchanged sparticle is
shown, together with the current experimental bounds, applicable in the 
case of  a
sparticle mass $\tilde m^2$ of 100 GeV. The bounds scale as $\tilde m^2$.
Dashes in the last column signify that no non-trivial bounds obtain from
low-energy FCNC processes. }
\normalsize
 
Another equally interesting possibility to see products of pairs of $\l$
couplings is to have sneutrino resonances in $\mu^+ \mu^- \rightarrow e^+
e^-~{\rm or}~\tau^+ \tau^-$. These will also arise from two
different $\lambda$-type couplings involving the same sneutrino. For these
signals, the SM background is large and it is again better to look at the
differential cross sections rather than the total rates, and use the same
isolation technique as for a single coupling. However, if it is possible
to tune the energy exactly to the mass of the resonance, then one can
certainly achieve very high accuracy in the measurement~\cite{Feng}. Some
of these products of $\lambda$ couplings could also lead to unlike-flavour
final states in $s$-channel processes. If both $s$- and $t$-channel 
processes are allowed, bounds on the relevant
products would closely resemble those on single couplings arising in
$\mu^+ \mu^- \rightarrow \mu^+ \mu^-$. Finally, an exciting possibility
arises when we consider $e\tau$ final states, since these can receive
contributions from the products $\lambda_{122} \lambda_{131},
\lambda_{232} \lambda_{133}$ and may help us to measure the couplings
$\lambda_{131},\lambda_{133}$, which are not measurable in isolation at a
muon collider.

In the upper half of Table 2, we have listed the possible products of
pairs of dissimilar $\lambda$ couplings that could be accessible at a muon
collider, together with the exchanged sparticle and the current
experimental bounds. Only 12 out of 36 possible products can be 
accessed\footnote{The remaining 24 can be found in Table 4.}.
As in Table 1, the (low-energy) bounds assume that the mass of the
exchanged sparticle is 100 GeV, with numbers in parentheses for the same
bounds when the exchanged sparticle has a mass of 1 TeV. 

Following the convention set up in Table 1, the lower half of Table 2
lists possible products of pairs of $\lambda'$ couplings that might be
investigated at a muon collider. Measuring these would require
improved tagging efficiencies, with the best chances being for 
$b \bar q$ final states. As in Table 1, we have listed four-jet states
arising from the production of single top quarks~\cite{singletop} for the 
sake of completeness, but we have not studied these states in this article.

It is also possible to investigate products of one $\lambda$-type coupling
with another one of the $\lambda'$ type. In this case, the $\lambda$
coupling appears in the coupling of a sneutrino resonance to a
muon--antimuon pair, while the $\lambda'$ coupling appears at the other
end, where a pair of quarks is produced.  For products of the form
$\lambda_{i22} \lambda'_{i33}$, one has the interesting possibility of a
(sneutrino) resonance contribution to scalar Higgs production at a muon
collider (which is not really surprising, since sneutrinos have the same
quantum numbers as Higgs bosons and can, in fact, mix with them if lepton
number is not conserved). This possibility has been studied in
Ref.~\cite{Feng} and will not be discussed further in this article.

In Fig.~4, we show the accessible part of the parameter space for the
cases when the final state consists of a fermion pair, and all tagging
efficiencies have been set to unity. The three cases correspond to 
exchange of a sneutrino in ($a$) the $s$-channel, with the possibility
of large resonance contributions, ($b$) the $t$-channel and ($c$) both
$s$- and $t$-channels, with smaller effects due to resonances.
As in Fig.~1, the dotted lines
correspond to bounds from LEP (on a different set of products, given in 
Table 4). We have
not shown the low-energy bounds on the products as there are too many of
them; however, they are listed in Table 2. To generate this figure, we
have assumed that a point in the parameter space is accessible if 5 or
more dissimilar fermion pairs are generated in 10~fb$^{-1}$ of data. This
criterion appears reasonable for leptons, but a larger number may be
necessary for jets. Of course, the curves will then simply scale as the
square root of that number. One must also scale the curves 
suitably by the relevant
tagging efficiencies. For example, for an $e\tau$ state, we need to
scale up the curves by $\sqrt{\epsilon_{e\tau}}$ where $\epsilon_{e\tau} =
\sqrt{\epsilon_{ee} \epsilon_{\tau\tau}} \simeq 0.85$. It is hardly
necessary to point out that better bounds can be obtained when there is a
$t$-channel exchange of a light sneutrino, since $s$-channel amplitudes
suffer a natural suppression because of the high energy of the FMC and the
NMC (unless a resonance is hit).

\begin{figure}[h]
\begin{center}
\vspace*{4.2in}
      \relax\noindent\hskip -4.2in\relax{\includegraphics{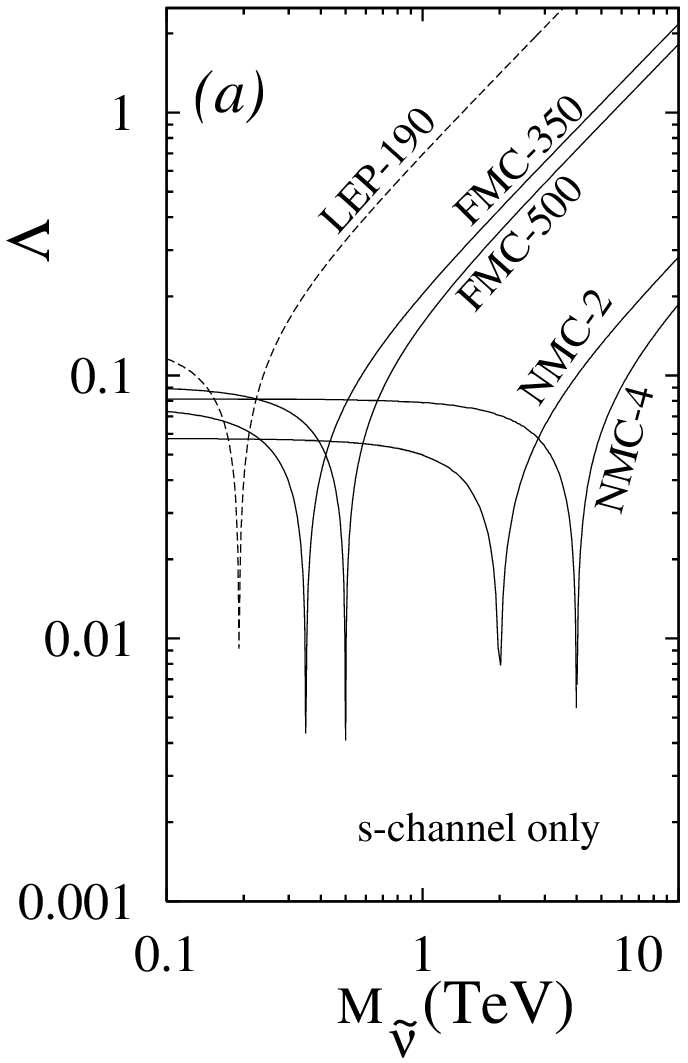}}
                     \hskip 1.99in\relax{\includegraphics{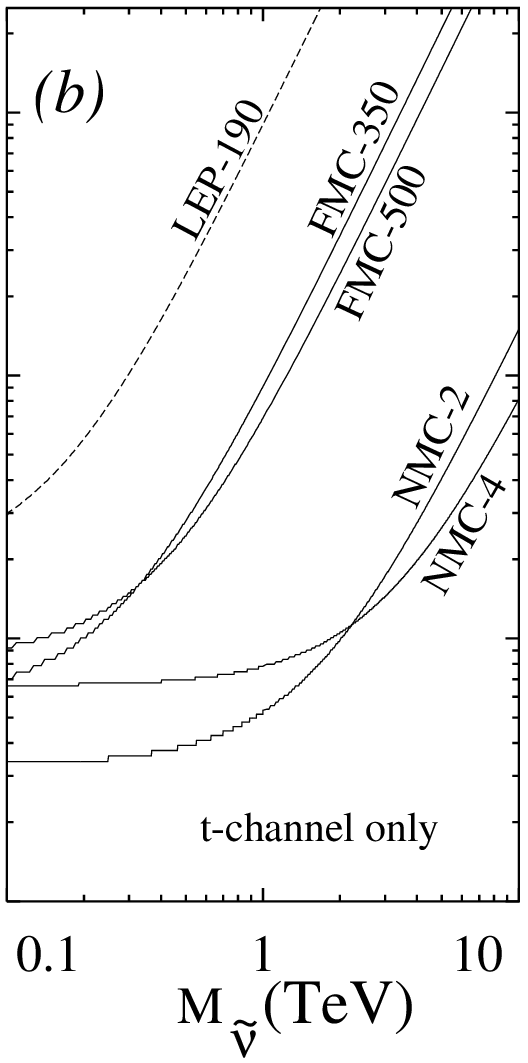}}
                     \hskip 2.005in\relax{\includegraphics{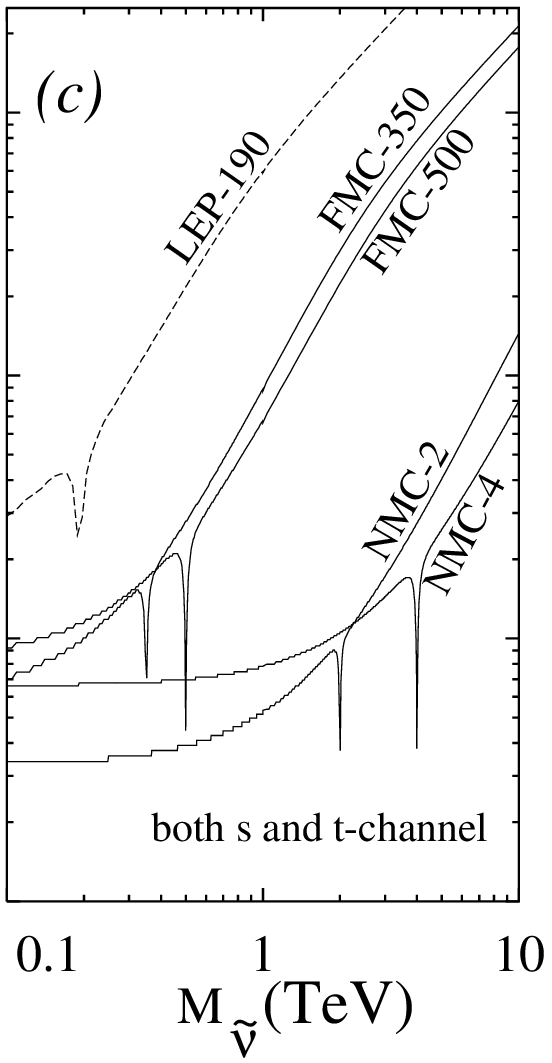}}
\end{center}
\vspace*{-1pt}
\caption{\footnotesize\em
Illustrating the reach of a muon collider in the \rp\ parameter space for
products $\Lambda$ of pairs of couplings when ($a$) an $s$-channel
sneutrino exchange is considered, ($b$) a $t$-channel sneutrino exchange
is considered, and ($c$) both $s$- and $t$-channel exchanges are
considered. Dotted lines showing the LEP bounds correspond to products
different from those accessible at a muon collider: they are only included for
comparison. All tagging efficiencies are set to unity.}
\end{figure}

For products of $\lp$ couplings, the exchanged particle will be a squark,
but the actual bounds will be very weak, since the number of excess events
will either be multiplied by the low $b$ and $c$ tagging efficiencies, or,
if one looks at untagged dijets, the excess must be greater than the 95\%
C.L. fluctuation in the expectation for the SM dijet production. The actual 
numbers for the latter, given the parameters chosen for this work, are about
196 for LEP, 1223 and 629 for the FMC at 350 GeV and 500 GeV respectively,
and 3268 and 911 for the NMC at 2 TeV and 4 TeV respectively. These large
values make it clear that it will not be possible to probe small values of
\rp\ couplings. We do not consider a study of products of two $\lp$ couplings
worth its while and shall not mention them any further in this 
article.

\section{$\!\!\!\!\!\!.$~~Comparison with an $e^+e^-$ Collider}

As explained in Section 2, the major motivation for building a muon
collider is the limited energy and luminosity of an $e^+e^-$ machine.
However, is is very likely that a 350--500 GeV linear $e^+e^-$ collider
will, in fact, be built. 
This is the so-called Next Linear Collider (NLC), for which
several plans exist.  Such a machine will probably have an integrated
luminosity of about 10~fb$^{-1}$, which makes it completely competitive
with the FMC, though falling short of both 
the energy and luminosity of the
NMC. Hence, as explained in Section 2, it is necessary to compare our
results at a muon collider with similar predictions at the NLC. The
physics analysis (within our assumptions and approximations) is rather
similar in the two cases. Even the angular cut of $20^\circ$ around the
beam pipe, which is imposed in the case of a muon collider, will be
required at the NLC to avoid beamstrahlung effects. Thus there is some
justification in using the numerical results we have obtained for the FMC
to make predictions for the NLC as well.

\vspace{-0.2in}
\footnotesize
\begin{center}
$$
\begin{array}{|c|c|c|c|c|}
\hline
 {\rm Final~state}
& {\rm Coupling~at}~\mu^+\mu^- & {\rm Exchange~at}~\mu^+\mu^-
& {\rm Coupling~at}~e^+e^- & {\rm Exchange~at}~e^+e^- \\
\hline
 e^+ e^-
 & \l_{121} & \snu_e   (t)   & \l_{121} & \snu_\mu (s+t)  \\
 & \l_{122} & \snu_\mu (t)   &*\l_{131} & \snu_\tau(s+t)  \\
 & \l_{132} & \snu_\tau(t)   &          &                \\
 & \l_{231} & \snu_\tau(t)   &          &                \\
\hline
\mu^+\mu^-
 & \l_{122} & \snu_e   (s+t) & \l_{121} & \snu_e   (t)  \\
 & \l_{232} & \snu_\tau(s+t) & \l_{122} & \snu_\mu (t)  \\
 &          &               & \l_{132} & \snu_\tau(t)  \\
 &          &               & \l_{231} & \snu_\tau(t)  \\
\hline
\tau^+ \tau^-
 & \l_{123} & \snu_e   (t)   & \l_{123} & \snu_\mu (t)  \\
 & \l_{132} & \snu_\tau(t)   & \l_{131} & \snu_e   (t)  \\
 &*\l_{232} & \snu_\mu (t)   &*\l_{133} & \snu_\tau(t)  \\
 &*\l_{233} & \snu_\tau(t)   & \l_{231} & \snu_\mu (t)  \\
\hline
\end{array}
$$
\end{center}
\noindent
{\normalsize\rm Table 3}. {\footnotesize\em
Comparison between a $\mu^+\mu^-$ and an $e^+e^-$ collider for
processes leading to bounds on individual \rp\ couplings of
the $LL\bar E$ type. An asterisk ($*$) indicates that the coupling
is measurable at only one of the two kinds of collider. }
\normalsize
 
The chief difference between an $e^+e^-$ collider and a $\mu^+\mu^-$
collider lies in the different flavour of the initial state, which means
that the same final state will produce bounds on an $R$-parity-violating
coupling with {\em different} flavour indices. 
To keep track of these, we list
in Table 3 the accessible couplings for the same final state at both
$e^+e^-$ and $\mu^+\mu^-$ colliders. We also mention (in parentheses)
whether an $s$- or a $t$-channel resonance is responsible for the excess
events predicted, which enables one to identify the relevant plots in
Figs.~1, 2 and 3. For example, for a single \rp\ coupling and an 
$s+t$-channel exchange, one must take the dashed (LEP) curve from 
Fig.~1($b$) and the solid (FMC) curves from Fig.~1($a$). For $t$-channel
exchanges, we need the dashed curve from Fig.~1($a$) and the solid
curves from Fig.~1($b$).

What immediately stands out from Table 3 is the fact that the
results from a $\mu^+\mu^-$ machine nicely {\em complement} those
obtainable at an $e^+e^-$ collider. The two couplings, $\l_{232}$ and
$\l_{233}$, which are not accessible at an $e^+e^-$ collider, are
accessible at a muon collider. On the other hand, the muon collider cannot
measure $\l_{131}$ and $\l_{133}$, which an $e^+e^-$ collider can. Clearly
it would be useful to have both sets of measurements. For the five couplings
that can be measured at both machines, the same complementarity holds. For
example, $\l_{121}$ is measurable at either machine, but it can be better
measured at a muon collider 

\vspace{-0.3in}
\footnotesize
\begin{center}
$$
\begin{array}{|c|c|c|c|c|c|}
\hline
{\rm Serial~No.} & {\rm Product}~\Lambda^2
& {\rm F.S.~at}~\mu^+\mu^- & {\rm Exchange~at}~\mu^+\mu^-
& {\rm F.S.~at}~e^+e^- & {\rm Exchange~at}~e^+ e^- \\
\hline
 1 & \l_{121} \l_{122} & e\mu     & \snu_e  ~ (s+t)
                       & e\mu     & \snu_\mu ~(s+t) \\
\hline
 2 & \l_{121} \l_{123} & e\tau    & \snu_e   ~(t)
                       & e\tau    & \snu_\mu ~(s+t) \\
\hline
 3,4,5 & \l_{121} \l_{13k} &    -     &    -
                       &    -     &        -       \\
\hline
 6 & \l_{121} \l_{231} &    -     &     -
                       & e\tau    & \snu_\mu ~(s) \\
\hline
 7 & \l_{121} \l_{232} &    -     &     -
                       & \mu\tau  & \snu_\mu ~(s) \\
\hline
 8 & \l_{121} \l_{233} &    -     &     -
                       & \tau\tau & \snu_\mu ~(s) \\
\hline
 9 & \l_{122} \l_{123} & \mu\tau  & \snu_e   ~(s+t)
                       & \mu\tau  & \snu_\mu ~(t) \\
\hline
10 & \l_{122} \l_{131} & e\tau    & \snu_e   ~(s)
                       &    -     &     -          \\
\hline
11 & \l_{122} \l_{132} & \mu\tau  & \snu_e   ~(s+t)
                       &    -     &     -          \\
\hline
12 & \l_{122} \l_{133} & \tau\tau & \snu_e   ~(s)
                       &    -     &     -          \\
\hline
13,14,15 & \l_{122} \l_{23k} &  -       &    -
                       &    -     &     -          \\
\hline
16,17,18 & \l_{123} \l_{13k} &  -       &    -
                       &    -     &     -          \\
\hline
19,20,21 & \l_{123} \l_{23k} &  -       &    -
                       &    -     &     -          \\
\hline
22 & \l_{131} \l_{132} &    -     &          -
                       & e\mu     & \snu_\tau~(s+t) \\
\hline
23 & \l_{131} \l_{133} &    -     &     -
                       & e\tau    & \snu_\tau~(s+t) \\
\hline
24 & \l_{131} \l_{231} &    -     &     -
                       & e\mu     & \snu_\tau~(s) \\
\hline
25 & \l_{131} \l_{232} & ee       & \snu_\tau~(s)
                       & \mu\mu   & \snu_\tau~(s) \\
\hline
26 & \l_{131} \l_{233} &    -     &     -
                       & \mu\tau  & \snu_\tau~(s) \\
\hline
27 & \l_{132} \l_{133} &    -     &     -
                       & \mu\tau  & \snu_\tau~(t) \\
\hline
28 & \l_{132} \l_{231} &    -     &     -
                       &    -     &     -          \\
\hline
29 & \l_{132} \l_{232} & e\mu     & \snu_\tau~(s+t)
                       &    -     &     -          \\
\hline
30 & \l_{132} \l_{233} &    -     &     -
                       &    -     &     -          \\
\hline
31 & \l_{133} \l_{231} &    -     &     -
                       &    -     &     -          \\
\hline
32 & \l_{133} \l_{232} & e\tau    & \snu_\tau~(s)
                       &    -     &     -          \\
\hline
33 & \l_{133} \l_{233} &    -     &     -
                       &    -     &     -          \\
\hline
34 & \l_{231} \l_{232} & e\mu     & \snu_\tau~(s+t)
                       &    -     &     -          \\
\hline
35 & \l_{231} \l_{233} & e\tau    & \snu_\tau~(t)
                       &    -     &     -          \\
\hline
36 & \l_{232} \l_{233} & \mu\tau  & \snu_\tau~(s+t)
                       &    -     &     -          \\
\hline
\end{array}
$$
\end{center}
\noindent
{\normalsize\rm Table 4}. {\footnotesize\em
Comparison between a $\mu^+\mu^-$ and an $e^+e^-$ collider for
processes leading to bounds on \rp\ products of the $LL\bar E$ type.
Dashes indicate inaccessibility, and $k = 1,2,3$ whenever it appears.}
\normalsize

\noindent
if muonic final states can be better identified, or at an $e^+e^-$ 
collider if it is the electronic
final states that can be better identified. 
One can also access different sneutrino resonances at the
two types of machine, which may lead to different cross sections if the
sneutrino masses are not degenerate.
 
For products of couplings, the graphs are very similar in the two cases,
but applicable to different sets of products.  We have noted that out of 36
products, only 12 can be accessed at a muon collider. Similarly 12
products can be
accessed at an $e^+e^-$ collider, with 4 products common to both. This
leaves only 16 products that cannot be bounded when both machines run.
This is illustrated in Table 4, where a full list of the products of
$\l$-type couplings is shown, illustrating accessibility at both
$e^+e^-$ and $\mu^+\mu^-$ colliders.

We see, then, that a muon collider, with the same energy and luminosity
parameters as an $e^+e^-$ collider, can yield important new information
about the nature of $R$-parity violation, especially in the case when the
sfermion masses are large and low-energy bounds on the couplings are
invalidated. With higher energies and luminosities, such as are planned
for the NMC, one can obviously probe a larger region of parameter space.
This provides at least one motivation for building such a machine.

\section{$\!\!\!\!\!\!.$~~Critique and Summary}

Before concluding, it would be well to reiterate the weak points of the
current analysis, which must be regarded as something of a zeroth-order
study of the problem. In the first place, the work is futuristic, since
nothing like a realistic design has been yet achieved for a muon collider,
though intensive studies are under way.  For this reason, energy and
luminosity parameters are assumed to be those given by the Accelerator
Physics Study Group at the Fermilab Workshop on the First Muon
Collider~\cite{FNAL}. These may change. It is also possible that the cuts
on the angular distribution made in our analysis are too weak/strong and a
future analysis can improve on it. The assumption of uniform detection
efficiencies and the exact values assumed will probably change
significantly, but though these will change the numerical results, none of
our conclusions will change qualitatively. Finally, our study of 
final-state jets was done assuming that a jet forms a cone around the 
direction of the parent
parton, and the thrust axis coincides with the parton momentum. This may
change a little on inclusion of fragmentation effects, although cumulative
experience with parton-level analyses suggests that such changes are
small. However, a more detailed study might change the bin width and hence
the number of bins, which is rather important for making the $\chi^2$
analysis envisaged here. To sum up, numbers will change here and there for
a variety of reasons, but we believe our general conclusions to be fairly
robust.

To summarize, then, we point out that the present constraints on some of
the \rp\ couplings are relatively weak. If the supersymmetric particles
are light enough to be produced copiously at LEP, we shall shortly be in
a position to witness dramatic signals. On the other hand, if they are
either too heavy or too weakly coupled to be produced --- as seems
increasingly likely --- indirect effects
provide us with a means to investigate this sector. While a high-energy
$e^+e^-$ collider such as the NLC is obviously one facility where such
searches can be carried out, we stress that a muon collider not only
complements these results, but carries some of them further if high
energies and luminosities can be attained.  Quite a few of the
lepton-number-violating \rp\ couplings lead to significant deviations in
the $\mu^+ \mu^- \rightarrow f \bar{f}$ angular distributions. For some of
the couplings, this effect can be used to impose bounds that are stronger
than any available today, while for others it will provide a complementary
test. This is particularly true if the sparticles turn out to have
masses close to or above a TeV. A similar analysis would also be
applicable to a large class of leptoquark and dilepton couplings. 
These results can, in fact, be mostly read-off from the ones presented 
in this article. Thus, it is clear that
studies of $R_p$-violating supersymmetry and related models would 
benefit considerabley from the construction of a muon collider.

\centerline{\bf Acknowledgements}

The authors are grateful to the organizers of the Workshop on the
First Muon Collider (Fermilab, 1997) for providing the motivation for
this work. They would also like to thank S.~Banerjee, J.~Feng and
N.K.~Mondal for discussions. SR acknowledges partial financial support
from the World Laboratory, Lausanne.

\newpage
\centerline{\large\bf Appendix}

Let us consider the process
\be  \dis
      \mu^-(p_1) + \mu^+(p_2) 
          \ra f(p_3) + \bar f(p_4) \ ,
\ee 
where $f$ is a generic fermion of mass $m_f$.  It is easy to ascertain
that the most general amplitude for such a process can be expressed as
\be
\barr{rcl}
   {\cal M} & = & \dis
           \sum_{a,b} \:
                \eta_{ab} \bar u(p_3) \gamma_\mu 
                                       P_a v(p_4) 
                \;        \bar v(p_2) \gamma^\mu      
                                       P_b u(p_1) 
            + \sum_{a,b} \: 
                \xi_{ab} \bar u(p_3) P_a v(p_4) 
                \;       \bar v(p_2) P_b u(p_1) 
          \\[1.5ex]
        & + & \dis
            \omega \; \bar u(p_3) \sigma_{\mu \nu} v(p_4) 
                \;       \bar v(p_2) \sigma^{\mu \nu} u(p_1) 
     \label{matrix-element}
\earr
\ee
where $a,b = L,R$ and $P_{L,R}$ are the chiral projection operators.  The
differential cross sections can then be expressed in terms of the
Mandelstam variables $s, t, u$ and the effective four-fermion coefficients
$\eta_{ab}$, $\xi_{ab}$ and $\omega$. For the sake of simplicity, we choose to
neglect the mass of the muon. On the other hand, we retain the effects of
possible muon polarization. The latter can be parametrized as
\be
\barr{rclcrcl}
   a_{LL} & = & (1 + \lambda_-) (1 + \lambda_+) 
          & \qquad &
   a_{LR} & = & (1 - \lambda_-) (1 + \lambda_+) 
         \\
   a_{RL} & = & (1 + \lambda_-) (1 - \lambda_+) 
          & \qquad &
   a_{RR} & = & (1 - \lambda_-) (1 - \lambda_+) 
        \label{polarization}
\earr
\ee
where $\lambda_{\pm}$ give the degree of left polarization for
$\mu^{\pm}$. We have then
\be
\barr{rcl}
\dis | {\cal M} |^2
     & = & \dis
         \left[ a_{LL} |\eta_{LL}|^2 + a_{RR} |\eta_{RR}|^2 \right]
           \;
         (u - m_f^2)^2 
              \\[1.5ex]
     & + & \dis
         \left[ a_{LL}  |\eta_{RL}|^2  + a_{RR} |\eta_{LR}|^2 \right]
           \; 
         (t - m_f^2)^2 
              \\[1.5ex]
     & + & \dis
         2 m_f^2 s 
         \left[ a_{LL} 
                 {\rm Re} \left(\eta_{LL} \eta_{RL}^\ast 
                          \right)
                + a_{RR} 
                 {\rm Re} \left(\eta_{RR} \eta_{LR}^\ast 
                          \right)
         \right]
               \\[1.5ex]
     & + & \dis
         \frac{s^2}{4} \;
         \left[ a_{LR} 
                 \left\{ |\xi_{RR} |^2 +  | \xi_{LR} |^2 
                  \right\}
              + a_{RL} 
                 \left\{ |\xi_{LL} |^2 +  | \xi_{RL} |^2 
                  \right\}
         \right]
               \\[1.5ex]
     & - & \dis
         \frac{m_f^2 s}{2} \;
         \left[ a_{LR} |\xi_{RR} + \xi_{LR} |^2 
              + a_{RL} |\xi_{LL} + \xi_{RL} |^2 
         \right]
               \\[1.5ex]
     & + & \dis
        8 | \omega|^2 (1 - \lambda_+ \lambda_-) \;
             \left[ (t - u)^2 + 2 m_f^2 s \right]
               \\[1.5ex]
     & + & \dis
        2 s (u - t)
         \left[ a_{LR} {\rm Re} (\omega^\ast \xi_{RR} ) 
              + a_{RL} {\rm Re} (\omega^\ast \xi_{LL} )
         \right]
       \ ,
\earr
\ee
and, finally,
\be
\frac{ {\rm d} \sigma}{ {\rm d} t} 
    = \frac{1}{16 \pi s^2} | {\cal M} |^2 N_c \,
\ee
$N_c$ being the appropriate colour factor.

The Standard Model expressions for $\eta_{ab}$ and $\xi_{ab}$ are
obviously determined in terms of the fermion couplings to the neutral
gauge bosons $V_i~(\equiv \gamma/Z)$. We choose to parametrize these as
\be
   \bar f \gamma^\mu 
        \left( \ell^{(f)}_i P_L + r^{(f)}_i P_R \right) \:f \; 
    V_i^{\mu} \,
\ee
with 
\be
\barr{rclcrcl}
r^{(f)}_1 & = & e Q_f  & \qquad &
                \ell^{(f)}_1 & = & e Q_f 
        \\[1.5ex]
r^{(f)}_2 & = & \dis \frac{e}{s_W c_W} (- s_W^2 Q_f)  & \qquad &
                \ell^{(f)}_1 & = & \dis \frac{e}{s_W c_W} (T_{3f} - s_W^2 Q_f)
         \ .
\earr
\ee
With the above definitions, the SM expressions (for $f \ne \nu_\mu$) are
given by
\be
\barr{rclcrcl}
\eta_{LL} & = & \dis \sum_i 
                 \left[ \frac{\ell^{(\mu)}_i \ell^{(f)}_i }
                                 { s - M_i^2 + i \Gamma_i M_i}
            +           \frac{ \left( \ell^{(\mu)}_i \right)^2 }
                                 { t - M_i^2} \delta_{\mu f}
                 \right]
          & \qquad &
\eta_{LR} & = & \dis \sum_i \frac{\ell^{(f)}_i r^{(\mu)}_i }
                                 { s - M_i^2 + i \Gamma_i M_i}
       \\[3ex]
\eta_{RR} & = & \dis \sum_i 
                 \left[ \frac{r^{(\mu)}_i r^{(f)}_i }
                                 { s - M_i^2 + i \Gamma_i M_i}
            +           \frac{ \left( r^{(\mu)}_i \right)^2 }
                                 { t - M_i^2} \delta_{\mu f}
                 \right]
          & \qquad &
\eta_{RL} & = & \dis \sum_i \frac{r^{(f)}_i \ell^{(\mu)}_i }
                                 { s - M_i^2 + i \Gamma_i M_i}
       \\[2ex]
\xi_{LL} & = & 0
          & \qquad &
\xi_{LR} & = & - 2 \dis \sum_i \frac{\ell^{(\mu)}_i r^{(\mu)}_i }
                                 { t - M_i^2} \delta_{\mu f}
       \\[2ex]
\xi_{RR} & = & 0
          & \qquad &
\xi_{RL} & = &\dis \xi_{LR}
        \\[2ex]
\omega & = & 0       & & & &
\earr
\ee

In the presence of $R_p$-violating interactions, some of the four-fermion
form-factors $\eta_{ab}$ and $\xi_{ab}$  will receive
corrections ($\omega$ remains identically zero), 
with the form depending on the particular final state. The
only non-zero corrections are:
\be
\barr{llrcl}
\mu^+ \mu^- \ra \mu^+ \mu^- &: \qquad&
     \Delta \eta_{LR} =  \Delta \eta_{RL} & = &
                \dis \sum_i 
                       \frac{ \lambda^2_{i22} }
                            { 2 (t - m^2_{\tilde \nu_i} ) }
            \\[2ex]
   & &  \Delta \xi_{LR} =  \Delta \xi_{RL} & = &
                \dis - \sum_i 
                       \frac{ \lambda^2_{i22} }
                            { 2 (s - m^2_{\tilde \nu_i} 
                                  + i \Gamma_{\tilde \nu_i}
                                       m_{\tilde \nu_i} ) }
        \\[3ex]
\mu^+ \mu^- \ra e_k^+ e_k^- \hspace*{1em}(k \ne 2)
              &: \qquad&
            \Delta \eta_{LR} & = &
                \dis \sum_i 
                       \frac{ \lambda^2_{ik2} }
                            { 2 (t - m^2_{\tilde \nu_i} ) }
            \\[2ex]
   & &  
            \Delta \eta_{RL} & = &
                \dis \sum_i 
                       \frac{ \lambda^2_{i2k} }
                            { 2 (t - m^2_{\tilde \nu_i} ) }
            \\[3ex]
\mu^+ \mu^- \ra d_k^+ d_k^- 
              &: \qquad&
            \Delta \eta_{RL} & = &
                \dis \sum_i 
                       \frac{ \lambda^{\prime 2}_{2ik} }
                            { 2 (t - m^2_{\tilde u_{iL} } ) }
\earr
\ee

\footnotesize


\newpage
\begin{thebibliography}{References}

\def\bib{\bibitem}
%
%
\def\tp{these proceedings}
\def\ib#1,#2,#3{       {\em ibid.\/ }{\bf #1} (19#2) #3}
\def\ap#1,#2,#3{       {\em Ann.~Phys.~(NY)\/ }{\bf #1} (19#2) #3}
\def\appb#1,#2,#3{     {\em Acta Phys.\ Polon.\/ }{\bf B#1} (19#2) #3}
\def\cpc#1,#2,#3{      {\em Comput. Phys. Commun.\/ }{\bf #1} (19#2) #3}
\def\ijmp#1,#2,#3{     {\em Int.~J.~Mod.~Phys.\/ } {\bf A#1} (19#2) #3}
\def\mpl#1,#2,#3 {     {\em Mod.~Phys.~Lett.\/ } {\bf A#1} (19#2) #3}
\def\np#1,#2,#3{       {\em Nucl.~Phys.\/ }{\bf B#1} (19#2) #3}
\def\npps#1,#2,#3{     {\em Nucl.~Phys.~B (Proc.~Suppl.)\/ }
                             {\bf B#1} (19#2) #3}
\def\plb#1,#2,#3{      {\em Phys.~Lett.\/ }{\bf B#1} (19#2) #3}
\def\pr#1,#2,#3{       {\em Phys.~Rev.\/ }{\bf #1} (19#2) #3}
\def\prd#1,#2,#3{      {\em Phys.~Rev.\/ }{\bf D#1} (19#2) #3}
\def\prep#1,#2,#3{     {\em Phys.~Rep.\/ }{\bf #1} (19#2) #3}
\def\prl#1,#2,#3{      {\em Phys.~Rev.~Lett.\/ }{\bf #1} (19#2) #3}
\def\prog#1,#2,#3{     {\em Prog.~Theor.~Phys.\/ }{\bf #1} (19#2) #3}
\def\rmp#1,#2,#3{      {\em Rev.~Mod.~Phys.\/ }{\bf #1} (19#2) #3}
\def\sp#1,#2,#3{       {\em Sov.~Phys.-Usp.\/ }{\bf #1} (19#2) #3}
\def\zpc#1,#2,#3{      {\em Z. Phys.\/ }{\bf C#1} (19#2) #3}

\bib{SM_revs} For recent reviews, see, for example, \\
W.~Hollik, Karlsruhe Univ. preprint KA-TP-22-1997 (1997),
hep-ph/9711492; \\
G.~Altarelli, CERN preprint CERN-TH-97-278 (1997), hep-ph/9710434. 

\bib{mssm} H.P.~Nilles, \prep110,84,{1};\\
  H.E.~Haber and G.L.~Kane, \prep117,85,{75}.

\bib{rpar} S.~Weinberg, \pr26,82,{287};\\
   N.~Sakai and T.~Yanagida, \np197,82,{533};\\
   C.S.~Aulakh and R.~Mohapatra, \plb119,82,{136}.

\bib{bilinear} M.A.~Diaz, Univ. of Valencia report no. IFIC-98-11 
(1998), hep-ph/9802407, and references therein; \\
   S.~Roy and B.~Mukhopadhyaya, \pr55,97,{7020}.

\bib{rpardef} G.~Farrar and P.~Fayet, \plb76,78,{575}.

\bib{rpar2}  F.~Zwirner, \plb132,83,{103};\\
  J.~Ellis \etal, \plb150,85,{142};\\
  G.G.~Ross and J.W.F.~Valle, \plb151,85,{375};\\
  S.~Dawson, \np261,85,{297};\\
  S. Dimopoulos and L.J.~Hall, \plb207,87,{210}.\\
  L.J. Hall, \mpl5,90,{467}.

\bib{HalSuz}  L.~J.~Hall and M.~Suzuki, \np231,84,{419}.

\bib{IbRo_92} L.E.~Iba\~nez and G.G.~Ross, \np368,92,{3}.

\bib{baryo}
A.~Bouquet and P.~Salati, \np284,87,557;\\
A.E.~Nelson and S.M.~Barr, \plb246,90,141;\\
B.A.~Campbell \etal, \plb256,91,457;\\
W.~Fischler \etal, \plb258,91,45.

\bib{DrRo_93} H.~Dreiner and G.G.~Ross, \np410,93,{188}.

\bib{Products} A.Yu.~Smirnov and F.~Vissani, \plb380,96,{317};\\
         K.~Agashe and M.~Graesser, \prd54,96,{4445};\\
         D.~Choudhury and  P.~Roy, \plb378,96,{153};\\
         C.E.~Carlson, P.~Roy and M.~Sher, \plb357,95,{94}.

\bib{BGH_89} V.~Barger, G.F.~Giudice and T.~Han, \pr40,89,{2987}.

\bib{BhCh_95} G.~Bhattacharyya and D.~Choudhury, \mpl10,95,{1699}.

\bib{GRT_93} R.~Godbole, P.~Roy and X.~Tata, \np401,93,{67}.

\bib{DoubleBeta}M.~Hirsch, H.V.~Klapdor-Kleingrothaus and 
           S.G.~Kovalenko, \prl75,95,{17};\\
        K.S.~Babu and R.N.~Mohapatra, \ib75,95,{2276}.

\bib{z-decay}G.~Bhattacharyya, J.~Ellis and K.~Sridhar, \mpl10,95,1583;\\
  G.~Bhattacharyya, D.~Choudhury and K.~Sridhar, \plb355,95,193.

\bib{Dreiner} H.~Dreiner, in `Perspectives on Supersymmetry', ed. G.L. Kane 
(World Scientific), hep-ph/9707435.

\bib{BKP_95} V.~Barger, W.-Y.~Keung and R.J.N.~Phillips, 
     \plb364,95,{27}; (E) \ib377,96,{486}.

\bib{lep2}LEP2 workshop proceedings (chap. {\em Search for New Particles})
(CERN, 1995).

\bib{Rp_at_Tevat}H.~Baer, C.~Kao and X.~Tata, \prd51,95,{2180} and
     references therein;\\
     M.~Guchait and D.P.~Roy, \prd54,96,{3276}; hep-ph/9707275.

\bib{HERA}
D.~Choudhury and S.~Raychaudhuri,
{\em Phys.~Lett.} {\bf B401} (1997) 54; \\
G.~Altarelli {\em et al.},
{\em Nucl.~Phys.} {\bf B506}, (1997) 3; \\
H.~Dreiner and P.~Morawitz,
{\em Nucl.~Phys.} {\bf B503}, (1997) 55; \\
J.~Kalinowski, {\em et al.}
{\em Z.~Phys.} {\bf C74} (1997) 595.

\bib{Drees} M.~Drees, {\em Phys.~Lett.} {\bf B403}, (1997) 353.
 
\bib{GhGoRa} D.K.~Ghosh, R.M.~Godbole and S.~Raychaudhuri,
in preparation.

\bib{Cheng} H.-C.~Cheng, Fermilab report no. FERMILAB-CONF-97-418-T
(1997), hep-ph/9712427.
 
\bib{Kajita} T.~Kajita, Talk delivered on behalf of the Kamiokande and
Super-Kamiokande Collaborations, Neutrino '98, Japan (1998),
http://www-sk.icrr.u-tokyo.ac.jp/nu98/scan/063/.
 
\bib{singletop} R.J.~Oakes {\em et al.}, \pr57,98,{534}; \\
A.~Datta {\em et al.},  \pr56,97,{3107}.
 
\bib{Feng} J.L.~Feng, J.F.~Gunion and T.~Han, Madison report No. UCD-97-25
 (1997),  hep-ph/9711414.

\bib{FNAL} See, for example, C.~Ankenbrandt, Fermilab report No.
FERMILAB-CONF-98-074 (1998), 
http://www-lib.fnal.gov/archive/1998/conf/Conf-98-074.html.

\bib{tariq}T.~Aziz (L3 Collaboration), private communication.

\bib{pdg}Particle Data Group, \pr50,94,1173.

\bib{mohap}R.~Mohapatra, \pr34,86,3457.

\end{thebibliography}
\end{document}